\documentclass[preprint,nofootinbib,amsmath,amsfonts,amssymb,onecolumn,superscriptaddress]{revtex4}
\usepackage{graphicx,psfrag,color}
\usepackage{bm}
\usepackage{mathbbol,verbatim}
\usepackage{slashed}
\usepackage{graphics}
\usepackage{multirow}
\allowdisplaybreaks
\usepackage{color,ulem}

\begin{document}

\title{Prospects for Triple Gauge Coupling Measurements at Future Lepton Colliders and the 14 TeV LHC}
\author{Ligong Bian}
\email{lgb@itp.ac.cn}
\affiliation{State Key Laboratory of Theoretical Physics and \\ Kavli Institute for Theoretical Physics China (KITPC), \\ Institute of Theoretical Physics, Chinese Academy of Sciences, Beijing 100190, P. R. China}
\author{Jing Shu}
\email{jshu@itp.ac.cn}
\affiliation{State Key Laboratory of Theoretical Physics and \\ Kavli Institute for Theoretical Physics China (KITPC), \\ Institute of Theoretical Physics, Chinese Academy of Sciences, Beijing 100190, P. R. China}
\affiliation{CAS Center for Excellence in Particle Physics, Beijing 100049, China}
\author{Yongchao Zhang}
\email{yczhang@pku.edu.cn}
\affiliation{State Key Laboratory of Theoretical Physics and \\ Kavli Institute for Theoretical Physics China (KITPC), \\ Institute of Theoretical Physics, Chinese Academy of Sciences, Beijing 100190, P. R. China}

\date{\today}

\begin{abstract}
  The $WW$ production is the primary channel to directly probe the triple gauge couplings. We first analyze the $e^+ e^- \rightarrow W^+ W^-$ process at the future lepton collider, China's proposed Circular Electron-Positron Collider (CEPC). We use the five kinematical angles in this process to constrain the anomalous triple gauge couplings and relevant dimension six operators at the CEPC up to the order of magnitude of $10^{-4}$. The most sensible information is obtained from the distributions of the production scattering angle and the decay azimuthal angles. We also estimate constraints at the 14 TeV LHC, with both 300 fb$^{-1}$ and 3000 fb$^{-1}$ integrated luminosity from the leading lepton $p_T$ and azimuthal angle difference $\Delta \phi_{ll}$ distributions in the di-lepton channel. The constrain is somewhat weaker, up to the order of magnitude of $10^{-3}$. The limits on the triple gauge couplings are complementary to those on the electroweak precision observables and Higgs couplings. Our results show that the gap between sensitivities of the electroweak and triple gauge boson precision can be significantly decreased to less than one order of magnitude at the 14 TeV LHC, and that both the two sensitivities can be further improved at the CEPC.
\end{abstract}
\maketitle

\tableofcontents
\newpage

\section{Introduction}

The observation of the standard model (SM) like Higgs boson at the large hadron collider (LHC) is a milestone of the elementary particle physics. In absence of any conclusive signal of new physics beyond SM, it is important in the forthcoming decades to pin down the electroweak (EW) symmetry breaking scenario, measure precisely the SM couplings and directly search for new physics at higher energy scales. All these could be closely related to the EW gauge sector in the SM. Precise determination of the gauge couplings is an essential part of high energy physics in the near future , e.g. to constrain the new physics from heavy states~\cite{Marzocca:2012zn}.

Due to non-Abelian nature of the weak interaction, there exist triple and quartic couplings among the EW gauge bosons in the SM. In this work we will focus on the charged triple gauge couplings (TGCs), i.e. those of form $WW\gamma$ and $WWZ$. The TGCs beyond SM can be parameterized in the framework of anomalous triple gauge couplings (aTGCs)~\cite{Hagiwara}, or in the language of effective field theory (EFT)~\cite{Weinberg1,Weinberg2,operator1,operator2,operator3}. In the sense of the phenomenological point of view, both the two scenarios are effective theories, valid only up to some specific scale, beyond which the unitarity of scattering amplitudes breaks down or the perturbation expansion (e.g. to dimension six order) does not make sense~\cite{unitarity1,unitarity2,unitarity3}. At the lowest order, it is straightforward to connect the Wilson coefficients to the anomalous couplings.

At $e^+ e^-$ colliders the TGCs can be directly probed in  the $WW$ pair, single-$W$ ($We\nu$) and single-photon ($\nu\nu\gamma$) processes. At hadron colliders, the di-boson final states $WW$, $WZ$ and $W\gamma$ can be used to study the charged gauge couplings. In the language of EFT, the Higgs-gauge couplings and oblique corrections are related to the TGCs, for instance, the universal gauge fermion coupling deviations can be re-shifted into $S-T$ and the triple gauge boson couplings \cite{Grojean}, thus the Higgs data and EW precision measurements can be used to constrain indirectly the gauge couplings~\cite{TGC:higgsdata}. In addition, the $WW\gamma$ coupling can induce rare processes at loop level such as $b \rightarrow s\gamma$, thus the observables from meson decays also contribute to constrain the aTGCs~\cite{TGC:Bmeson}.

Direct measurements of the charged TGCs has been implemented on LEP, Tevatron and LHC~\cite{LEP,CDF,D0,D02,D03,ATLAS1,ATLAS2,ATLAS3,ATLAS4,ATLAS5,CMS1,CMS2,CMS3,CMS4}, and the current most stringent bounds are mainly from the $W$ pair measurements at LEP II~\cite{LEP}, with the aTGCs at the order of magnitude of few times $10^{-2}$~\cite{aTGC:comparison1,aTGC:comparison2}. It is expected that the sensitivities would be improved by one to two orders of magnitudes at the International Linear Collider~\cite{ILC:TDR}, due to the larger luminosity, higher energy and polarized beams. Very recently a proposal for an alternative future $e^+ e^-$ collider has been mad in China, the Circular Electron-Positron Collider (CEPC)~\cite{CEPC:preCDR}. Part of this work is devoted to estimations of the constraints on aTGCs at CEPC in the $W$ pair channel.

Taking into account the decays of $W$ bosons, the process $e^+e^- \rightarrow W^+W^-$ can be described by five kinematic angles, one ($\cos\theta$) for $W$ production and the rest four for the decay products~\cite{Hagiwara,Bilenky}. We use the differential cross sections with respect to the five angles to set limits on the TGCs at CEPC. Though only $\cos\theta$ depends directly on the TGCs, we find that the four decay angles also contribute substantially in constraining the gauge couplings, which however depends largely on the $W$ decay channels and the aTGCs involved. As a whole, the sensitivities at CEPC can reach up to the order of magnitude of $10^{-3}$ to $10^{-4}$, comparable to that at ILC~\cite{ILC:TDR} or ever better. The $WW$ process at hadron colliders is somewhat similar to that at lepton colliders; for the former the dominate channel at parton level is $q\bar{q} \rightarrow W^+W^-$. We estimate also the constraints at the 14 TeV LHC with a luminosity of 300 fb$^{-1}$ or 3000 fb$^{-1}$, which can largely improve the current bounds.

The rest of this work is organized as follows: In the next section we set up the framework for discussing the process $e^+e^- \rightarrow W^+W^-$, where we clarify the aTGCs and Dim-6 operators involved, conventions for the five angles, etc. In section III, we study analytically the response of the differential cross sections to the aTGCs and show them graphically. From these plots one can judge qualitatively which differential distributions are more sensitive to the aTGCs and which aTGC can be more severely constrained. Section IV is devoted to estimations of the sensitivities of aTGCs and the Dim-6 operators involved at CEPC,  with a center-of-mass energy of $\sqrt{s} = 240$ GeV and an integrated luminosity of 5 ab$^{-1}$. In this section,  we also show explicitly the separate contributions of sensitivities due to the five production and decay angles. In section V we shed some light on the TGCs at hadron colliders and estimate na\"ively the bounds on the TGCs at 14 TeV LHC. The present and future constraints on the aTGCs from lepton and hadron colliders are collected at the end of this section. In section VI, We comment briefly on the complementarity of the direct TGC measurements and the indirect constraints coming from EW precision data and Higgs data, before we conclude in the last section.

\section{Preliminaries}
\subsection{Anomalous triple gauge couplings beyond the SM}

With the anomalous contributions beyond SM~\cite{Hagiwara}, the charged TGCs among the SM EW gauge bosons can be generally parameterized as,
\begin{eqnarray}
\label{eqn:lagrangianTGC}
{\cal L}_{\rm TGC}/g_{WWV} &=&
ig_{1,V} \Big( W^+_{\mu\nu}W^-_{\mu}V_{\nu} -W^-_{\mu\nu}W^+_{\mu}V_{\nu} \Big)
+ i\kappa_V W^+_\mu W^-_\nu V_{\mu\nu}
+ \frac{i\lambda_V}{M_W^2} W^+_{\lambda\mu} W^-_{\mu\nu} V_{\nu\lambda} \nonumber \\
&& + g_5^V \varepsilon_{\mu\nu\rho\sigma} \Big( W^+_{\mu} \overleftrightarrow{\partial}_\rho W_\nu \Big) V_\sigma
- g_4^V W^+_\mu W^-_\nu \Big( \partial_\mu V_\nu + \partial_\nu V_\mu \Big) \nonumber \\
&& + i \tilde{\kappa}_V W^+_\mu W^-_\nu \tilde{V}_{\mu\nu}
+ \frac{i \tilde{\lambda}_V}{M_W^2} W^+_{\lambda\mu} W^-_{\mu\nu} \tilde{V}_{\nu\lambda} \,.
\end{eqnarray}
where $V = \gamma\,, Z$, the gauge couplings $g_{WW\gamma} = -e$, $g_{WWZ} = -e\cot\theta_W$ with $\cos\theta_W$ being the weak mixing angle, the field strength tensor $F_{\mu\nu} \equiv \partial_\mu A_\nu - \partial_\nu A_\mu$ with $A = W\,, \gamma\,, Z$, and the conjugate tensor $\tilde{V}_{\mu\nu} = \frac12 \varepsilon_{\mu\nu\rho\sigma} V_{\rho\sigma}$, and $A \overleftrightarrow{\partial_\mu} B \equiv A ({\partial_\mu} B) - ({\partial_\mu} A) B$. Besides the SM TGCs, the Lagrangian Eq.~(\ref{eqn:lagrangianTGC}) contains 14 anomalous TGCs up to dimension six in the most general form. The couplings $g_{1,\,V}$, $\kappa_V$ and $\lambda_V$ are both parity ($P$) and charge conjugate ($C$) conserving and the rest eight are $C$ or $P$ violating. In the SM, $g_{1,V} = \kappa_V = 1$ whereas all others are vanishing. Electromagnetic gauge symmetry requires that $g_{1,\,\gamma} = 1$ and $g_{4,\,\gamma} = g_{5,\,\gamma} = 0$. Consequently, in the absence of C or P violation from beyond SM physics, there are only five aTGCs:
\begin{eqnarray}
\label{eqn:aTGC5}
\Delta g_{1,Z}\,, \quad \Delta\kappa_\gamma\,,\quad \Delta \kappa_Z\,, \quad \lambda_\gamma\,, \quad \lambda_Z \,,
\end{eqnarray}
where we have split the SM and new contributions apart, $\Delta g_{1,Z} \equiv g_{1,Z} - 1$, $\Delta \kappa_{\gamma,\,Z} \equiv \kappa_{\gamma,\,Z} - 1$.

If the beyond SM physics is described in the language of EFT, then only a few dimension-6 operators are relevant to the charged ($C$ and $P$ conserving) TGCs, in the SILH basis~\cite{SILH,SILH2},
\begin{eqnarray}
\label{Lagrangian}
\Delta {\cal L} 
&=& \frac{i c_W\, g}{2M_W^2}\left( H^\dagger  \sigma^i \overleftrightarrow {D^\mu} H \right )( D^\nu  W_{\mu \nu})^i
+\frac{i c_{HW} \, g}{M_W^2}\, (D^\mu H)^\dagger \sigma^i (D^\nu H)W_{\mu \nu}^i \nonumber \\
&& +\frac{i c_{HB}\, g^\prime}{M_W^2}\, (D^\mu H)^\dagger (D^\nu H)B_{\mu \nu} \,
+\frac{c_{3W}\,  g}{6 M_W^2}\, \epsilon^{ijk} W_{\mu}^{i\, \nu} W_{\nu}^{j\, \rho} W_{\rho}^{k\, \mu} \,.
\end{eqnarray}
In this convention, the term $\frac{c_{WB}}{M_W^2} B^{\mu\nu}W^i_{\mu\nu} (H^\dagger \sigma^i H)$ can be expressed as a linear combination of other terms by integration by parts. The $c_W$ operator contribute to the oblique parameter $S$~\cite{oblique}, and is tightly constrained by EW precision measurements, $\sim 10^{-5}$~\cite{LEP,ILC:TDR,EWPT,LTWang}. The $W$ pair bound on $c_{W}$ at ILC and CEPC is only of the order of magnitude of $10^{-4}$. Thus as a first order approximation, we can neglect the $c_W$ term, with only three operators left at the Dim-6 level which are related to the aTGCs via~\cite{operator1,operator2,operator3}
\begin{eqnarray}
\label{eqn:tgc2operator}
\Delta  g_{1,Z} &=&  - c_{HW} / \cos^2\theta_W  \,, \nonumber \\
\Delta \kappa_\gamma &=&  - ( c_{HW} + c_{HB} ) \,, \nonumber \\
\lambda_\gamma &=&  - c_{3W} \, ,
\end{eqnarray}
and $\Delta \kappa_Z = \Delta g_{1,Z} -  \tan^2\theta_W \Delta \kappa_\gamma$,  $\lambda_\gamma = \lambda_Z$. Under such circumstance, the aTGCs are related by the EW $SU(2)_L \times U(1)_Y$ gauge symmetry, and there is only three independent couplings in the $C$ and $P$ conserving sector, e.g. the three explicitly given in the equation above. Given any constraints on the aTGCs from the present and future high energy colliders, we can always translate them into limits on the relevant dimension-6 operators and apply them to any particular models in connection with charged triple gauge couplings.

\subsection{$W$ pair production at $e^+ e^-$ colliders}

At tree level, the $e^+ e^- \rightarrow W^+ W^-$ process are mediated by $s$-channel $\gamma$/$Z$ and $t$-channel neutrino. In the most general case, oblique corrections, non-standard gauge-fermion couplings and aTGCs all contribute to the $W$ pair cross section. Due to the severe constraints from EW precision measurements, it is a good approximation to neglect the corrections from the oblique terms and beyond SM gauge-fermion interactions and focus only on the effects of aTGCs~\cite{Falkowski}.\footnote{In~\cite{Buchalla}, it is argued that in the language of EFT the gauge fermion interactions in $e^+ e^- \rightarrow W^+ W^-$ are related to the TGCs by redefining the gauge fields. In light of this, it is more reasonable to say we are working in a basis without the extra gauge fermion interactions. See also~\cite{Grojean}.} At $\sqrt{s} = 240$ GeV, the designed energy for CEPC, the production cross section is dominated by the neutrino mediated transverse $WW$ configuration $(-+)$, which forms a peak in the forward region $\cos\theta \sim 1$. As this helicity state does not depend on any TGCs, it is the dominate irreducible background for measuring the aTGCs, especially when the colliding energy goes higher. Thus precise determination of the TGCs requires a large statistics at lepton colliders.

When the information of $W$ boson decay is taken into consideration, the kinematics of $e^+ e^- \rightarrow W^+ W^- \rightarrow f_1 \bar{f}_2 \bar{f}_3 f_4$ is dictated by five angles in the narrow $W$ width approximation~\cite{Hagiwara,Bilenky,Bilenky2,ambiguities}: the scattering angle $\theta$ between $e^-$ and $W^-$, the polar angles $\theta^\ast_{1,2}$ and the azimuthal angles $\phi^\ast_{1,2}$ of the down-type (anti-)fermion in the rest frame of $W^\mp$. To unambiguously define these decay angles, we define the right-handed coordinate systems of the W rest frames such that the $z$ axis is along the $W^{\mp}$ flight direction $\overrightarrow{W^\mp}$, and $y$ in the direction of $\overrightarrow{e^-} \times \overrightarrow{W^\mp}$, where $\overrightarrow{e^-}$ is the direction of electron beam. The five-fold differential cross section reads~\cite{Bilenky,Bilenky2,Gounaris}
\begin{eqnarray}
\label{eqn:diff}
\frac{{\rm d}\sigma (e^+e^- \rightarrow W^+W^- \rightarrow f_1 \bar{f}_2 \bar{f}_3 f_4)}
{ {\rm d}\cos\theta {\rm d}\cos\theta^\ast_1 {\rm d}\phi^\ast_1 {\rm d}\cos\theta^\ast_2 {\rm d}\phi^\ast_2 }
&=& {\rm BR} \cdot \frac{\beta}{32\pi s} \left( \frac{3}{8\pi} \right)^2
\sum_{\lambda\tau_1\tau'_1\tau_2\tau'_2}  F^{(\lambda)}_{\tau_1\tau_2} F^{(\lambda)\ast}_{\tau'_1\tau'_2} \nonumber \\
&& \times D_{\tau_1 \tau'_1} (\theta^\ast_1,\phi^\ast_1) D_{\tau_2 \tau'_2} (\pi-\theta^\ast_2,\pi+\phi^\ast_2) \,,
\end{eqnarray}
with the branching ratio ${\rm BR} = {\rm Br} (W^- \rightarrow \bar{f}_3 f_4) {\rm Br} (W^+ \rightarrow f_1 \bar{f}_2)$, $\beta$ being the velocity of $W$ boson, $F$ and $D$ being the helicity amplitudes for $WW$ production and the $W$ decay matrix, $\lambda$ and $\tau^{(\prime)}$ being the helicities of $e^-$ and $W^\mp$. Integrating out some of the angles, we can obtain the the more inclusive differential cross sections,
\begin{eqnarray}
\label{eqn:distributions}
\frac{{\rm d}\sigma}{ {\rm d}\cos\theta} \,, \;\;
\frac{{\rm d}\sigma}{ {\rm d}\cos\theta^\ast_1} \,, \;\;
\frac{{\rm d}\sigma}{ {\rm d}\phi^\ast_1} \,, \;\;
\frac{{\rm d}\sigma}{ {\rm d}\cos\theta^\ast_2} \,, \;\;
\frac{{\rm d}\sigma}{ {\rm d}\phi^\ast_2} \,,
\end{eqnarray}
which can be extracted from experimental data, at least in principle. The polarization of $W$ bosons can also be described by the spin density matrix (SDM), with the two-particle joint SDM defined as~\cite{Bilenky}
\begin{eqnarray}
\label{eqn:SDM}
\rho_{\tau_1\tau'_1\tau_2\tau'_2} (s,\cos\theta) \equiv
\frac{ \sum_{\lambda} F^{(\lambda)}_{\tau_1\tau_2} F^{(\lambda)\ast}_{\tau'_1\tau'_2} }
{ \sum_{\lambda\tau_1\tau_2} \left| F^{(\lambda)}_{\tau_1\tau_2} \right|^2 } \,,
\end{eqnarray}
which is normalized to unity. The SDM has $3^4-1=80$ independent elements, and contains the full helicity information of the $W$ pairs~\cite{Gounaris}.

\section{Differential distributions in presence of the anomalous couplings}
\label{sec:distribution}

In presence of aTGCs, both the total and (five-fold or more inclusive) differential cross sections are affected. One may use the SDM elements or their appropriate linear combinations obtainable from experiments to constrain the anomalous couplings~\cite{Hagiwara,Bilenky,ILC:TGC2}. We resort alternatively to the differential cross sections in Eq.~(\ref{eqn:distributions}) with regard to the five kinematic angles, which are more physically intuitive. It is sometimes more convenient to replace the polar ($\cos\theta_{1,\,2}^\ast$) and azimuthal angles ($\phi_{1,\,2}^\ast$) by the corresponding angles for the quark jets ($\jmath$) and charged leptons ($\ell$): $\cos\theta^\ast_{\jmath,\,\ell}$ and $\phi^\ast_{\jmath,\,\ell}$. Then the differential cross sections reads
\begin{eqnarray}
\label{eqn:distributions2}
\frac{{\rm d}\sigma}{ {\rm d}\cos\theta} \,, \;\;
\frac{{\rm d}\sigma}{ {\rm d}\cos\theta^\ast_\ell} \,, \;\;
\frac{{\rm d}\sigma}{ {\rm d}\phi^\ast_\ell} \,, \;\;
\frac{{\rm d}\sigma}{ {\rm d}\cos\theta^\ast_\jmath} \,, \;\;
\frac{{\rm d}\sigma}{ {\rm d}\phi^\ast_\jmath} \,.
\end{eqnarray}
It is worth emphasizing that only the five-fold differential cross section Eq.~(\ref{eqn:diff}) (and the SDM Eq.~(\ref{eqn:SDM})) contains the full information of $WW$ production and decay; considering only the five inclusive distributions above would lose some sensitivity, e.g. those from spin correlations between the $W$ pairs. However, the lost sensitivities are expected to be small~\cite{ILC:TGC2}. As we will show below, in addition to $\cos\theta$, distributions of the decay angles $\cos\theta^\ast_{\ell,\,\jmath}$ and $\phi^\ast_{\ell,\,\jmath}$ contribute significantly to the sensitivities.

The total cross section has a dependence on the aTGCs $\alpha_i$ in the quadratic form\footnote{Both the total cross section and angular distributions have quadratic dependence on the aTGCs, and there exists a two-fold ambiguity in obtaining sensitivities of these aTGCs, one close to the SM scenario, and the other one much farther away. This ambiguity can be removed by combining the total and differential cross sections~\cite{ambiguities}.},
\begin{eqnarray}
\sigma_{\rm total} = \sigma_0 \left( 1 + b_i \alpha_i + b_{ij}\alpha_i\alpha_j \right)
\end{eqnarray}
where $b_i$ and $b_{ij}$ are nominal linear and quadratic coefficients and $\sigma_0$ is the SM cross section. When the couplings $\alpha_i$ are sufficiently large, the quadratic term will eventually overcome the linear one and the total cross section is larger than in the SM (requiring that the coefficients $b_{ii}$ are positive definite, at least theoretically). However, when the couplings are very small, e.g. in the range of $10^{-4}$ to $10^{-3}$ of interest for the prospects at CEPC, we find that the linear terms always dominate over the quadratic ones $b_i\alpha_i \gg b_{ij} \alpha_i \alpha_j$, as expected. This is a reasonable consequence, as with such small aTGCs the higher order contributions from new physics beyond SM is always neglectable.
At the center-of-mass energy $\sqrt{s}=240$ GeV for CEPC, we have the SM cross section $\sigma_0=17.2$ pb. Given the convention in Eq.~(\ref{eqn:lagrangianTGC}), all the $b_i$ for the three $C$ and $P$ conserving couplings in Eq.~(\ref{eqn:tgc2operator}) are negative (at lower energies it is also possible that some of the $b_i$ are positive)\footnote{It is preferable to discuss the aTGCs in Eq.~(\ref{eqn:tgc2operator}) which respect the EW gauge symmetry. The main results, e.g. the $b_i$ and CEPC constraints, for the five most general $C$ and $P$ conserving aTGCs in Eq.~(\ref{eqn:aTGC5}) are collected in Appendix A.}
\begin{alignat}{2}
\label{eqn:bi}
b_1 &= b (\Delta g_{1,Z}) &= -0.120 \,, \nonumber \\
b_2 &= b (\Delta \kappa_{\gamma}) &= -0.155 \,, \nonumber \\
b_3 &= b (\lambda_{\gamma}) &= -0.104 \,.
\end{alignat}
which means that at leading order the total cross section decreases in presence of a positive aTGC. These coefficients are at the same order, which implies that the constraints of total cross section on these aTGCs are close to each other.\footnote{The $b_{i\, L,\,R}$ for different initial electron helicities $e^-_{L,\,R}$ are
\begin{alignat}{2}
& b_{1L} = -0.247 \,,   \quad && b_{1R} = + 0.127  \,, \nonumber \\
& b_{2L} = -0.0642 \,,  \quad && b_{2R} = - 0.0909  \,, \nonumber \\
& b_{3L} = -0.111 \,,   \quad && b_{3R} = + 0.00730  \,. \nonumber
\end{alignat}
Though in SM the $W$ pair production cross section for initial $e^-_R$ is almost 100 times smaller than that for $e^-_L$, the leading order corrections from aTGCs can be of the same order, and they can have either the same or opposite signs.}
We can write Eq.~(\ref{eqn:tgc2operator}) in a matrix form $\alpha_i = V_{ij} c_i$, with the rotating matrix
\begin{alignat}{2}
V = \left( \begin{matrix}
-\cos^{-2}\theta_W & 0 & 0 \\
-1 & -1 & 0 \\
0 & 0 & -1
\end{matrix} \right) \,.
\end{alignat}
It is straightforward to translate the linear coefficients $b_i$ to that for the Dim-6 operators, $b'_i = V_{ji} b_j$:
\begin{alignat}{2}
b'_1 &= b' (c_{HW}) &= +0.309 \,, \nonumber \\
b'_2 &= b' (c_{HB}) &= +0.155 \,, \nonumber \\
b'_3 &= b' (c_{3W}) &= +0.104 \,.
\end{alignat}
In the basis of EFT, all the $b_i$ at $\sqrt{s} = 240$ GeV are positive. The arguments here are based only on measurements of the total cross section of $W$ pair production; when distributions of the five kinematic angles are considered, more information of $WW$ production and decay is used, and the sensitivities are expected to be largely improved.

To examine the response of the angular distributions to the aTGCs, analogue to the case for the total cross section, we expand the differential cross sections Eq.~(\ref{eqn:distributions}) in terms of the aTGCs,
\begin{eqnarray}
\label{eqn:omegai}
\frac{{\rm d}\sigma}{{\rm d}\Omega_k} = \frac{{\rm d}\sigma_0}{{\rm d}\Omega_k} \Big[ 1 + \omega_i (\Omega_i) \alpha_i + \omega_{ij}  (\Omega_k) \alpha_i\alpha_j \Big] \,,
\end{eqnarray}
where $\Omega = \cos\theta$, $\cos\theta^\ast_{1,\,2}$, $\phi^\ast_{1,\,2}$ (or alternatively $\Omega = \cos\theta$, $\cos\theta^\ast_{\ell,\,\jmath}$, $\phi^\ast_{\ell,\,\jmath}$). Once again for sufficiently small aTGCs we can safely neglect the quadratic terms. It is straightforward to obtain analytically the linear coefficient functions $\omega_i$ from the differential cross sections\footnote{The expansion of differential cross sections in terms of the aTGCs and the $\omega$ coefficients are used by the LEP experimental groups~\cite{ALEPH,DELPHI,L3,OPAL} to extract constraints on the anomalous couplings~\cite{Diehl:1993br}.}, Which are collectively given in Appendix~\ref{sec:abi}. As for the $b_i$ coefficients, the distribution functions for the three relevant Dim-6 operators can be expressed as $\omega'_i = V_{ji} \omega_j$. The numerical functions $\omega_i (\Omega_k)$ for the aTGCs in Eqs.~(\ref{eqn:aTGC5}) and (\ref{eqn:tgc2operator}) are collected in Appendix~\ref{sec:bi}. Integrating over the angles $\frac{1}{\sigma_0}\int {\rm d} \Omega_k \, \omega_{i} (\Omega_k) \frac{{\rm d}\sigma_0}{{\rm d}\Omega_k}$ recovers the linear coefficients $b_{i}$, thus these functions $\omega_{i} (\Omega_k)$ measure the ``angular distributions'' of the coefficients $b_{i}$. In other words, the $\omega_i$ functions are the angular distributions of the aTGC effects. Combing the distributions $\omega_i (\Omega_k)$ and the differential cross sections in the SM $d\sigma_0/d\Omega_k$, one can easily identify, for one specific $\alpha_i$,  which angle is the most sensitive and even which part of the corresponding distribution deviate most from SM. Likewise one can judge qualitatively for one specific angle $\Omega_k$ which coupling $\alpha_i$ induces the largest deviation and thus is most stringently limited.

\begin{figure}[t]
  \centering
  \includegraphics[width=0.31\textwidth]{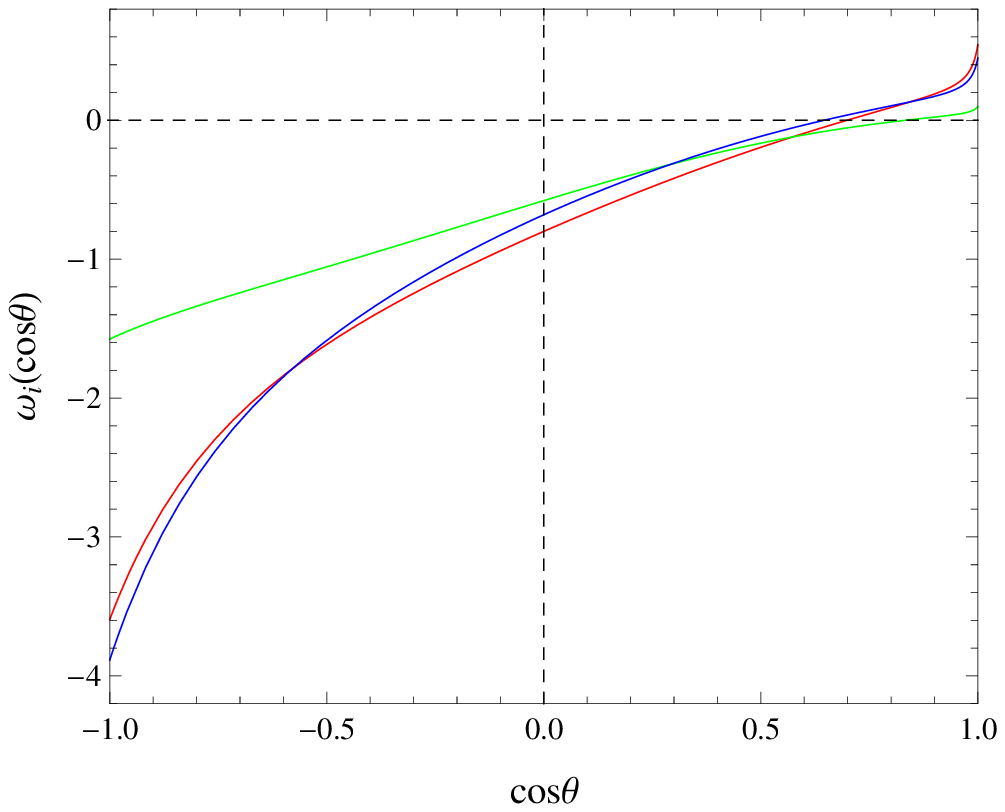}
  \includegraphics[width=0.315\textwidth]{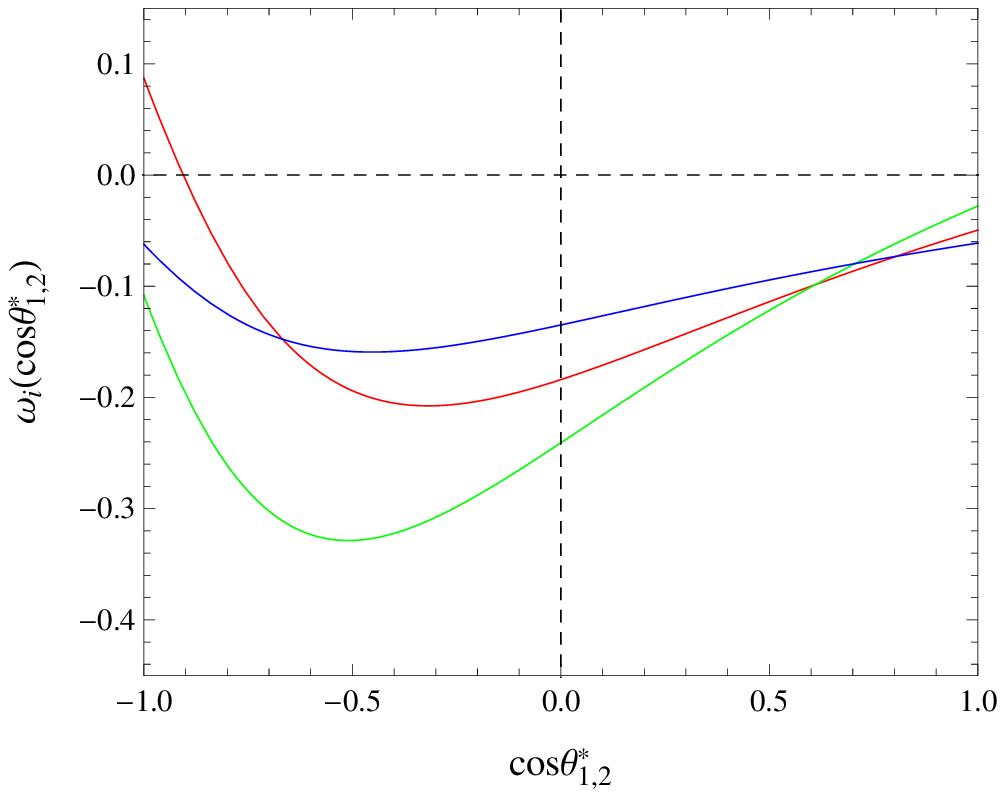}
  \includegraphics[width=0.31\textwidth]{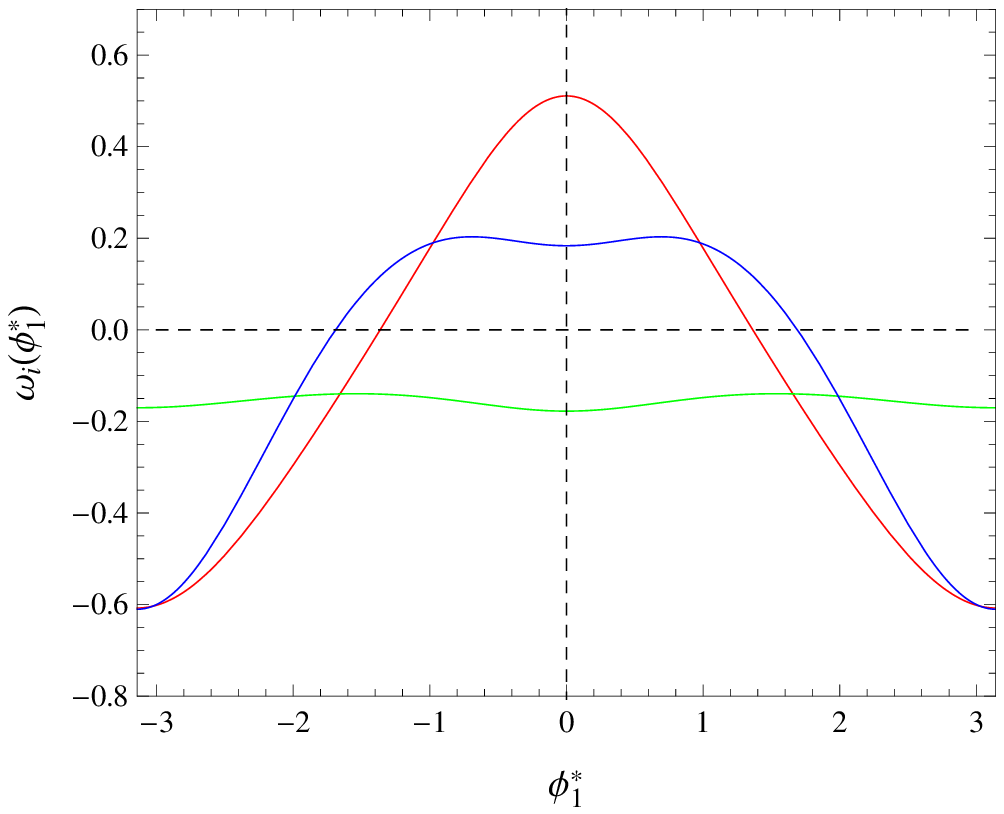} \\
  \includegraphics[width=0.31\textwidth]{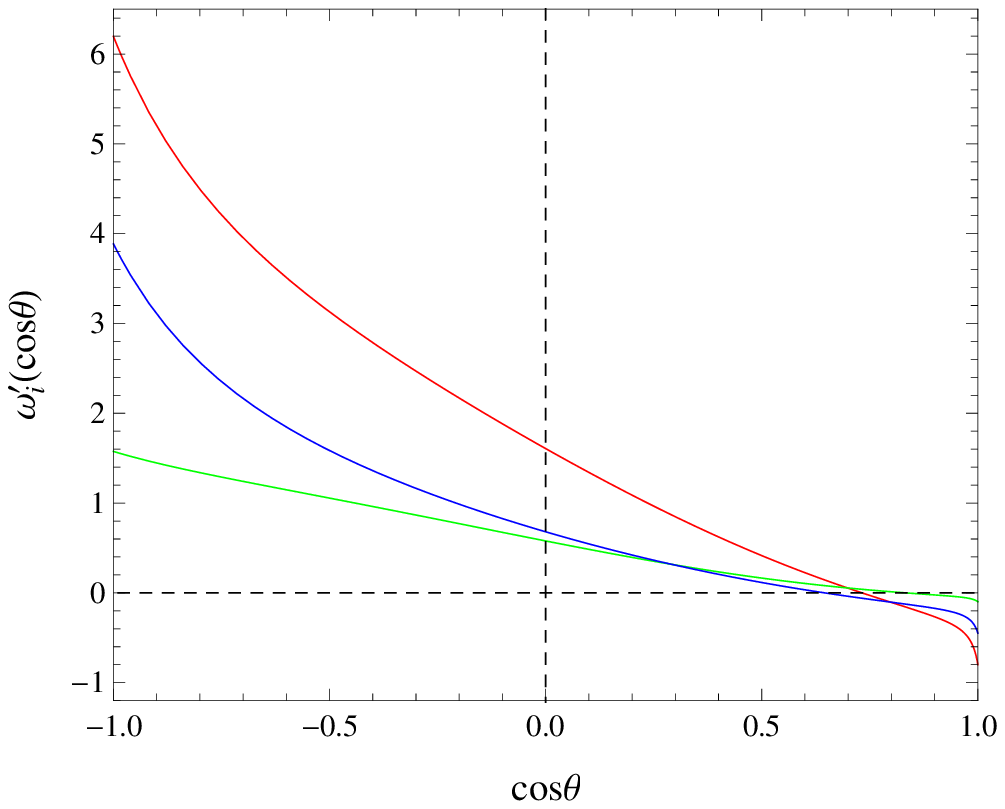}
  \includegraphics[width=0.315\textwidth]{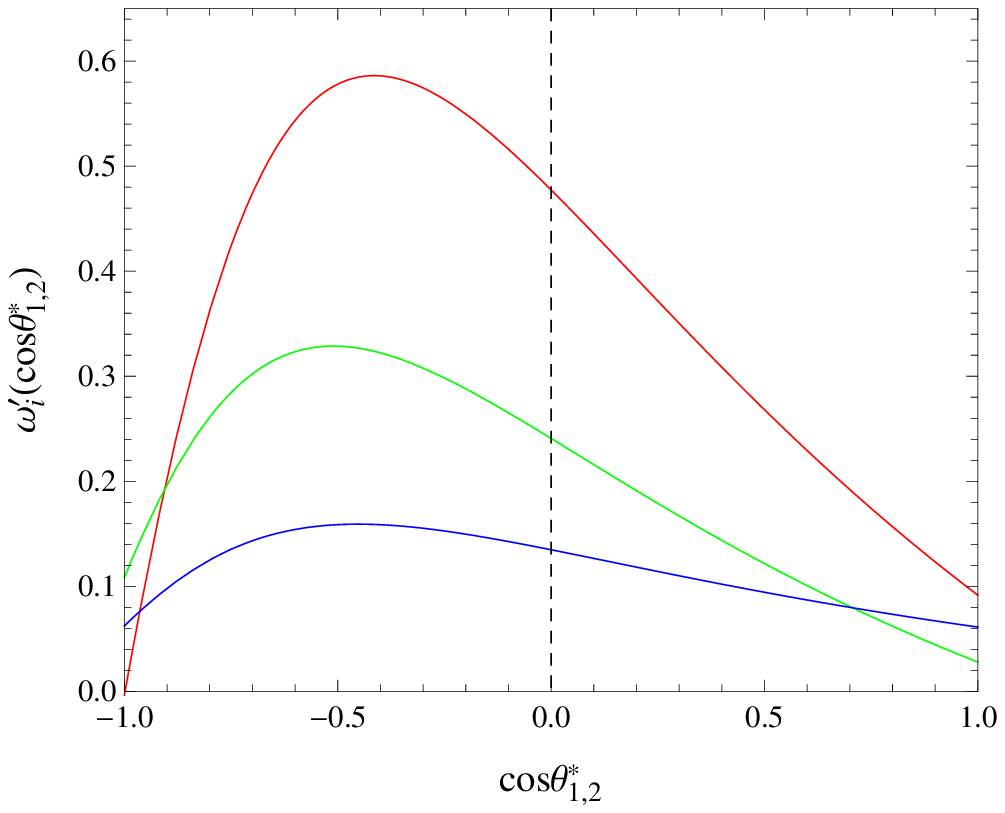}
  \includegraphics[width=0.31\textwidth]{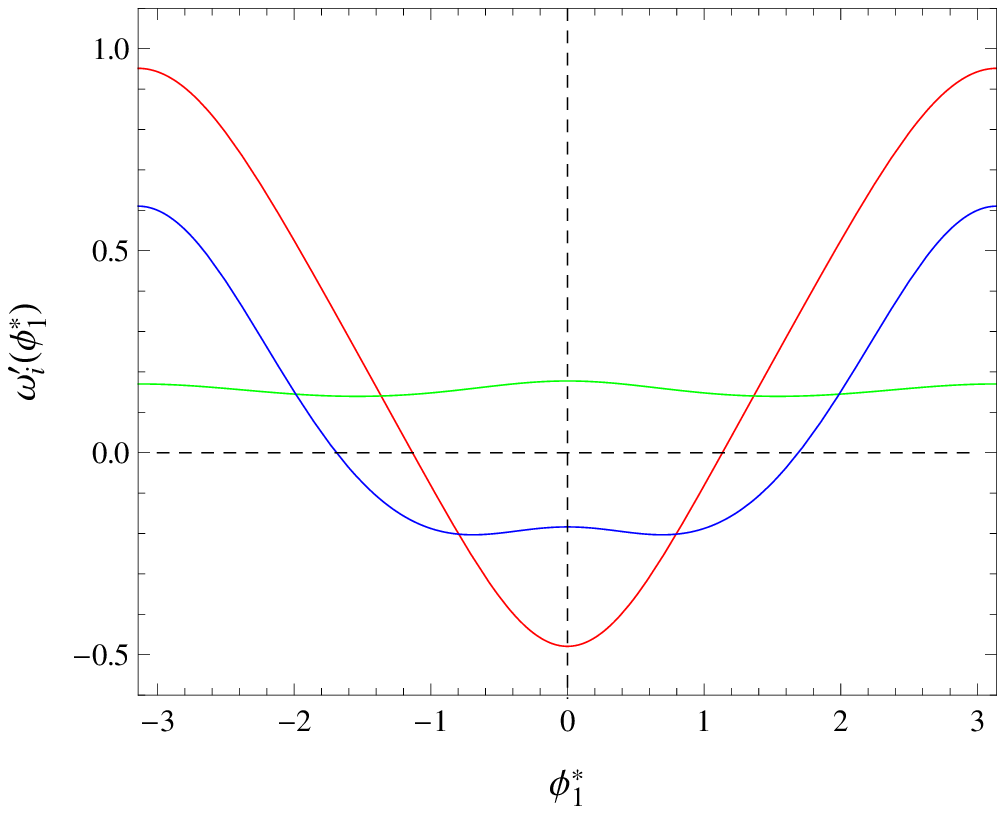}
  \vspace{-.5cm}
  \caption{Distributions $\omega_i (\Omega_k)$ as functions of the five angles $\Omega_k = \cos\theta$, $\cos\theta^\ast_{1,\,2}$, $\phi^\ast_{1,\,2}$, for the three aTGCs $\alpha_i = \Delta g_{1,Z}$, $\Delta \kappa_{\gamma}$ and $\lambda_{\gamma}$ in Eq.~(\ref{eqn:tgc2operator}) (and the three Dim-6 operators $c_{HW}$, $c_{HB}$ and $c_{3W}$ in the lower panels) denoted respectively by the red, green and blue curves. The center-of-mass energy for all the plots are $\sqrt{s} = 240$ GeV. See text for details. }
  \label{fig:omega}
\end{figure}

The coefficient functions $\omega_i (\Omega_i)$ for the three aTGCs and Dim-6 operators in Eq.~(\ref{eqn:tgc2operator}) are depicted in Fig.~\ref{fig:omega}, where we set the energy $\sqrt{s} = 240$ GeV. Due to hermiticity of weak interaction, distributions for the two polar angles $\omega_i (\cos\theta^\ast_{1,\,2})$ are identical, and $\omega_i (\phi^\ast_1)$ only differs from $\omega_i (\phi^\ast_2)$ by shifting a phase of $\pi$: $\phi^\ast_1 + \pi \rightarrow \phi^\ast_2$. Among the five angles, only the scattering angle $\cos\theta$ depends directly on the TGCs, and the sensitivities from the decay angles are mainly due to their correlation to the angle $\cos\theta$. Thus it is a natural consequence that $\cos\theta$ is generally most sensitive to the anomalous couplings. One can also arrive at such qualitative feature by comparing the magnitudes of $\omega_i$ for all the five angles in Fig.~\ref{fig:omega}: a larger deviation generally means a larger sensitivity.

It is transparent in the panel for $\cos\theta$ that, for all the three $C$ and $P$ conserving aTGCs in Eq.~(\ref{eqn:tgc2operator}), the largest deviation occurs in the backward region $\cos\theta\sim-1$. However, the sensitivity in this region of parameter space is somehow limited by the small cross sections, or equivalently say the number of events, especially for high energy collisions. Therefore, a huge statistics, e.g. on CEPC, can largely enhance the sensitivities in the backward region. In contrast, the deviation from SM in the forward direction is much smaller. However, due to the much larger cross section, the huge statistics in the forward region could also contribute substantially to the sensitivities. In addition, the $W$ pair events far away form the backward and forward regions are less affected by the aTGCs and contribute less to the sensitivities.

It is expected that as a whole the polar angle distributions contribute least to the sensitivities, as the magnitudes of the three curves in the middle panel of Fig.~\ref{fig:omega} are the smallest. Among these three curves, the magnitudes of $\omega_{} (\cos\theta^\ast_{1\,2})$ for $\Delta \kappa_\gamma$ is significantly larger than other couplings, implying that the polar angles are most sensitive to the couplings $\Delta \kappa_\gamma$ and contribute significantly in constraining it. Taking into account the SM distributions ${\rm d} \sigma_0 / {\rm d} \cos\theta^\ast_{1,\,2}$, the plot implies further that $\cos\theta^\ast_{1,\,2} \sim 0$ is the most sensitive region in the $W$ rest frame.
For the azimuthal angles $\phi^\ast_{1,\,2}$, we can easily see in the right panels that for $\Delta g_{1,\,Z}$ and $\lambda_\gamma$ the significant deviations occurs at $\phi^\ast_{1,\,2} \sim 0 ,\, \pm\pi$ in the rest frame of $W$ bosons. Thus it is expected that distributions of the azimuthal angles suffer significant distortions due to these non-vanishing aTGCs, which in turn severely constrains the couplings and improves largely the sensitivities. For $\Delta\kappa_\gamma$ the function $\omega_{}(\phi^\ast_{1,\,2})$ is almost a constant $\sim-0.17$, which means that given a non-vanishing $\Delta\kappa_\gamma$ the distributions ${\rm d}\sigma/{\rm d}\phi^\ast_{1,\,2}$ are rescaled by a factor of $(1-0.17\Delta\kappa_\gamma)$.

\section{CEPC constraints}

\subsection{Constraints on the TGCs and Dim-6 operators}

With a huge luminosity of 5 ab$^{-1}$ at CEPC at $\sqrt{s} = 240$ GeV, we can collect a total number of $8.6\times 10^7$ events of $W$ pairs, with 45\%, 44\% and 11\% decaying respectively in the hadronic, semileptonic and leptonic channels. With such a huge statistics, these anomalous TGCs are expected to be severely constrained. In this section we use the differential cross sections with respect to the five angels for $WW$ production and decay, i.e. Eq.~(\ref{eqn:distributions2}), to extract the generator level constraints on the $C$ and $P$ conserving aTGCs and relevant Dim-6 operators. It is fortunate that the radiative corrections, reducible backgrounds from non-$WW$ processes and systematic errors are small and well understood, especially for the non-hadronic decays~\cite{ILC:TGC1,ILC:TGC2}. We first estimate the statistical errors for these aTGCs in all the three distinctive channels (and combining all the available channels together) and then comment briefly on the corrections and systematic errors. In these channels not all the five angles can be fully reconstructed unambiguously, and it is unavoidably that we would loose some sensitivities for the TGCs. Here follow some comments on the ambiguities:
\begin{itemize}
  \item Due to the large branching ratio and high reconstruction efficiency, the semileptonic decays are the optimal channel. In this case, the charge of $W$ boson is assigned from the lepton charge\footnote{The $\tau$ lepton from $W$ decay is highly boosted, thus the $\tau\nu jj$ events are also viable to reconstruct the $W$ leptonic decays, as did in the LEP experiment~\cite{ALEPH,DELPHI,L3,OPAL}, although its reconstruction is much more complicated than the electrons or muons.}, and the only ambiguity is from the hadronic decay where one can not distinguish the quark jets from the antiquarks. We choose the jets from $W$ decay in the region $0 \leq \phi^\ast \leq \pi$. In other words, only the symmetric part of the $D$ decay function under the transformation $(\theta^\ast,\, \phi^\ast) \leftrightarrow (\pi-\theta^\ast,\, \pi+\phi^\ast)$ is used.
  \item The events in the hadronic channel appear to be four jets on colliders. Assuming the four jets can be correctly paired and the signs of $W$ bosons are correctly assigned, there leaves only the ambiguity in assigning jets into quarks and antiquarks, as in the semileptonic case.
  \item For the leptonic channel, we consider only the $e$ and $\mu$ channels as the events involving $\tau$ leptons can not be fully reconstructed due to the extra neutrinos from $\tau$ decay. In the limit of vanishing $W$ width, there is a two-fold ambiguity in solving the momenta of neutrino to reconstruct the $W$ bosons~\cite{Hagiwara}. Assuming the physical solution can be clearly distinguished from the unphysical one, we can then reconstruct fully the $W$ pair events in the purely leptonic final states.
\end{itemize}

We split simply distributions of the five angles $\cos\theta$, $\cos\theta^\ast_{1,\,2}$ and $\phi^\ast_{1,\,2}$ into a number of bins, count the event numbers in each bin, and use the shape of these distributions to set limits on the aTGCs. For large numbers of events, which is indeed the case for CEPC, the statistical errors can be estimated to be $\sqrt{N_i}$ with $N_i$ the number of events in the $i$th bin. It is then straightforward to define the standard $\chi^2$,
\begin{eqnarray}
\chi^2 \equiv \sum_{i} \left( \frac{N_i^{\rm bSM} - N_i^{\rm SM}}{\sqrt{N_i^{\rm SM}}} \right)^2 \,,
\end{eqnarray}
where $N_i^{\rm bSM}$ and $N_i^{\rm SM}$ are, respectively, the numbers of events in the $i$th bin for some specific distributions in the presence of beyond SM interactions and in the SM. Estimations of the one-parameter limits on the aTGCs and the relevant Dim-6 operators in Eq.~(\ref{eqn:tgc2operator}) are presented in Table~\ref{tab:sensitivity}, where all other anomalous couplings or Dim-6 coefficients are fixed to zero. In this table and the calculations below, all the distributions are split evenly into ten bins\footnote{We calculate also the sensitivities using respectively 20 and 50 bins and find that the sensitivities can only be slightly improved, at most by $\sim2\%$. In realistic analysis, given the data sets, the binning of events are chosen to optimize the sensitivities.}, and the ambiguity for hadronic decays has been taken into consideration. As stated above, the constraints are so strong that the quadratic terms of the aTGCs in the differential cross sections can hardly lead to any substantial effect on the sensitivities. Here are some comments on the constraints:
\begin{itemize}
  \item It is transparent that in the leptonic channel the limits on the aTGCs and Dim-6 operators are of the order of magnitude of few times $10^{-4}$ to $10^{-3}$. Due to the larger branching ratios, the semileptonic and hadronic channels can improve the constraints by a factor of two or three. When all the three channels are combined together, i.e. the last row ``all'' in Table~\ref{tab:sensitivity}, the constraints are even stronger and can reach close to the order of magnitude of $10^{-4}$.
  \item Dictated by the relation connecting the aTGCs and Dim-6 operators, the constraints on $c_{HW}$ combines those on the anomalous couplings $\Delta g_{1,\,Z}$ and $\Delta \kappa_{\gamma}$, and it is more severely constrained than $c_{HB}$ and $c_{3W}$. With regard to the coefficients $c_{HB} \sim - \kappa_{\gamma}$ and $c_{3W} \sim -\lambda_\gamma$, it is expected that at linear level the limits on $c_{HB}$ and $\Delta\kappa_\gamma$ (and $c_{3W}$ and $\lambda_\gamma$) should be the same. The tiny differences in Table~\ref{tab:sensitivity} are due to the quadratic corrections.
  \item Instead of fixing $\sqrt{s} = 240$ GeV, we consider also an alternative energy scan mode for CEPC, much like the LEP II~\cite{LEP}, keeping the mean energy at 240 GeV and the accumulated luminosity equal to 5 ab$^{-1}$. For instance, we can set $\sqrt{s} = 220$ to 260 GeV, with the step size being 5 GeV. Although constraints on the aTGCs get stronger when the energy goes higher, the energy scan mode gets much similar sensitivities on the aTGCs and Dim-6 operators to the running solely at 240 GeV, and does not help to improve the limits.
\end{itemize}

\begin{table}[t]
  \begin{center}
  \caption{Estimations of the $1\sigma$ prospects (in units of $10^{-4}$) for the aTGCs and Dim-6 operators in Eq.~(\ref{eqn:tgc2operator}) in the leptonic, semileptonic and hadronic decay channels of $WW$ process at CEPC with $\sqrt{s} = 240$ GeV and an integrated luminosity of 5 ab$^{-1}$. The last row ``all'' combines all the available channels above. All the sensitivities in this table are one parameter constraint where all other couplings or coefficients are fixed to zero. See text for details.}
  \label{tab:sensitivity}
  \begin{tabular}{cccccccc}
  \hline\hline
  channels & $\Delta g_{1,Z}$ & $\Delta \kappa_\gamma$ &  $\lambda_\gamma$ &$\;\;$& $c_{HW}$ &  $c_{HB}$ & $c_{3W}$ \\ \hline
  leptonic     &  5.90 & 9.87 & 6.57 &&  3.36 & 9.91 & 6.58 \\ \hline
  semileptonic &  2.19 & 3.33 & 2.35 &&  1.18 & 3.34 & 2.35 \\ \hline
  hadronic     &  2.51 & 3.37 & 2.54  &&  1.26 & 3.37 & 2.54 \\ \hline
  all          &  1.59 & 2.30 & 1.67 &&  0.84 & 2.31 & 1.67 \\
  \hline\hline
  \end{tabular}
  \end{center}
\end{table}

\begin{figure}[t]
  \centering
  \includegraphics[width=0.31\textwidth]{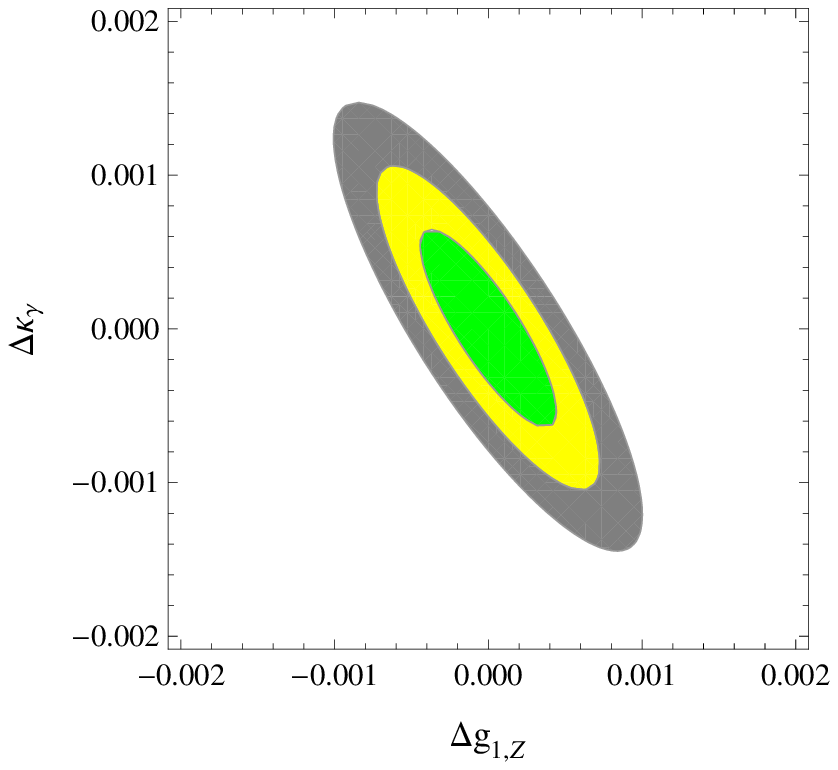}
  \includegraphics[width=0.31\textwidth]{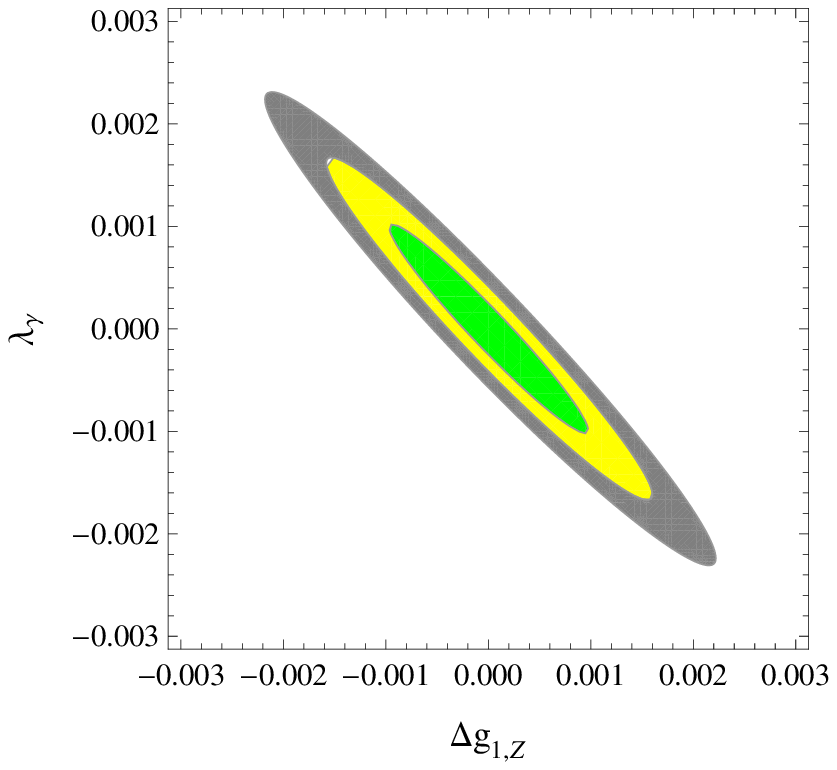}
  \includegraphics[width=0.32\textwidth]{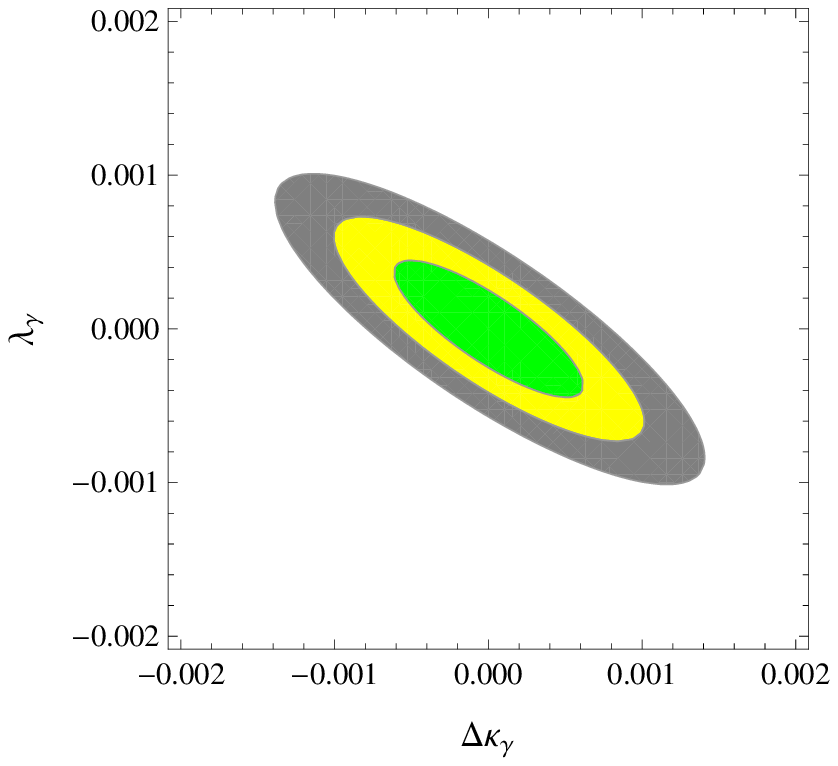} \\
  \includegraphics[width=0.31\textwidth]{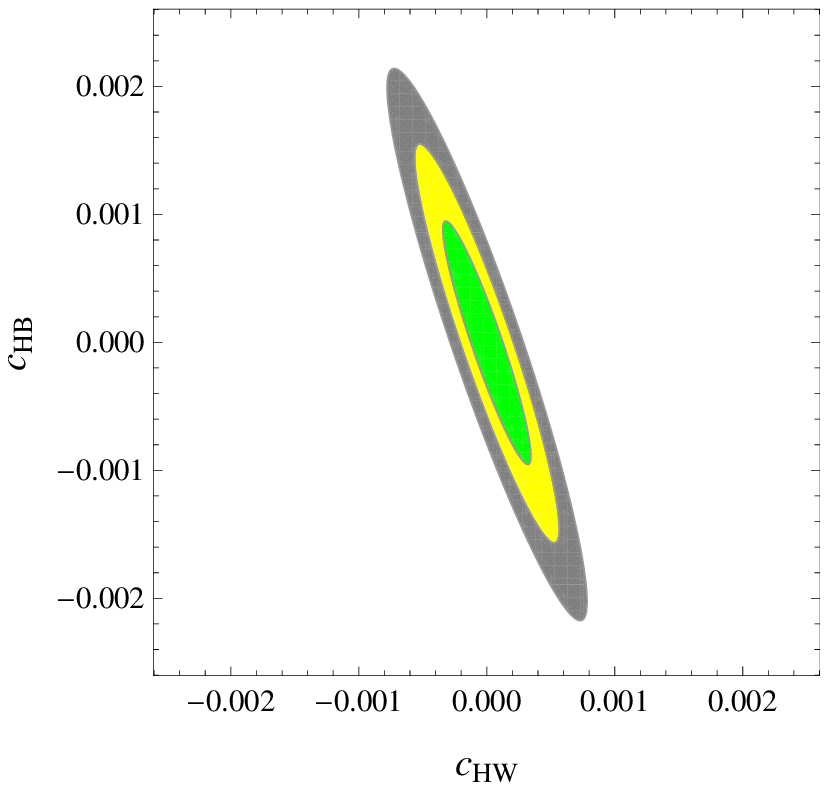}
  \includegraphics[width=0.31\textwidth]{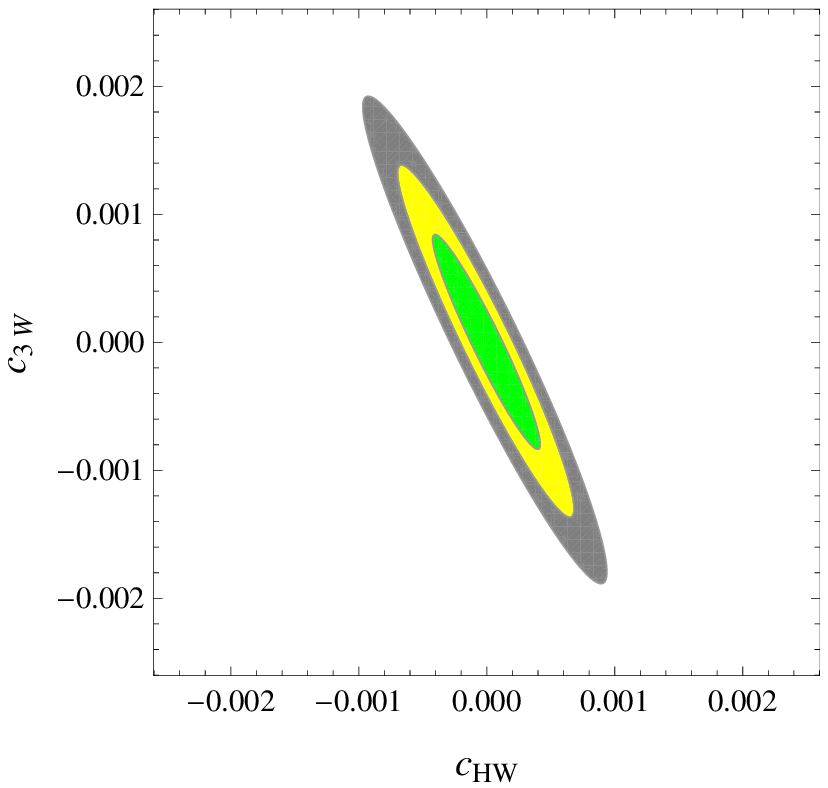}
  \includegraphics[width=0.32\textwidth]{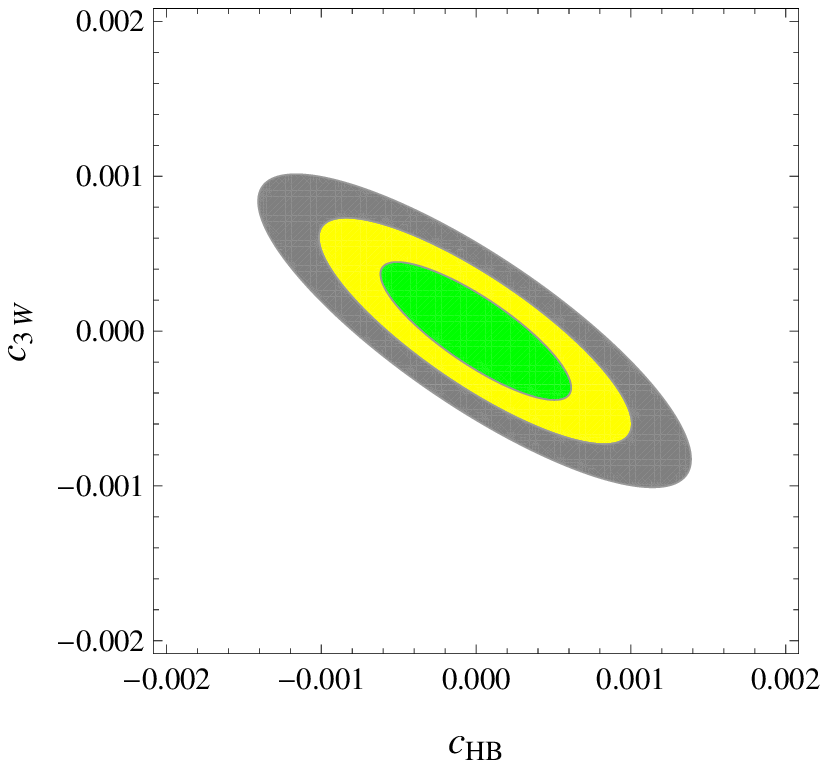}
  \vspace{-.5cm}
  \caption{1$\sigma$ (green), $2\sigma$ (yellow) and $3\sigma$ (gray) allowed regions for the couplings $\Delta g_{1,Z}$, $\Delta \kappa_\gamma$, $\lambda_\gamma$ (upper panels) and coefficients $c_{HW}$, $c_{HB}$, $c_{3W}$ (lower panels) at CEPC with $\sqrt{s} = 240$ GeV and a luminosity of 5 ab$^{-1}$ (combining all the available channels). In drawing the plots, two of the three couplings are allowed to vary and the third one is fixed to zero.}
  \label{fig:contours}
\end{figure}

The correlations between the three aTGCs and Dim-6 operators are, respectively,
\begin{eqnarray}
\label{eqn:rho}
\rho^{\rm aTGC} = \left( \begin{matrix}
  1 && \\
  0.839 &  1 &  \\
  0.969 &  0.824 &  1
\end{matrix} \right) \,, \quad
\rho^{\rm operator} = \left( \begin{matrix}
  1 && \\
  0.930 &  1 & \\
  0.954 &  0.824 &  1
\end{matrix} \right) \,,
\end{eqnarray}
when we have combined all the three decay channels. We also calculate the two-parameter constraints on the anomalous couplings and Dim-6 operators, with two of the three couplings or coefficients being allowed to vary and the third one being fixed to zero. The $1\sigma$, $2\sigma$ and $3\sigma$ regions for the three couplings $\Delta g_{1,Z}$, $\Delta \kappa_\gamma$, $\lambda_\gamma$ and coefficients $c_{HW}$, $c_{HB}$, $c_{3W}$ are presented in Fig.~\ref{fig:contours}, where we use all the available decay channels. In obtaining the contours, we set the standard $\Delta\chi^2$ values for two independent variables: for $1\sigma$, $2\sigma$ and $3\sigma$ errors $\Delta\chi^2$ equals 2.30, 6.18 and 11.83 respectively.

We can read from Eq.~(\ref{eqn:rho}) and Fig.~\ref{fig:contours} that some of the anmalous couplings are strongly correlated, such as $\Delta g_{1,Z}$ and $\lambda_\gamma$, and the direction $\Delta g_{1,Z} + \lambda_\gamma$ is much less severely constrained than in other combinations. However, we would like to stress that the correlations of the couplings, no matter whether these couplings are related by some symmetries, depend both on the theoretical predictions and the experimental data. In presence of the future CEPC data, correlations between the anomalous couplings might be dramatically changed. By the way, the potential blind directions might be removed by incorporating the helicity information of $e^\pm$ and $W^\pm$~\cite{blinddirection,Falkowski}.

Before turning to the next subsection we comment briefly on the experimental effects and the systematic errors. The dominate reducible backgrounds are the non-$WW$ four-fermion processes and the $q\bar{q}$ two-fermion process~\cite{ALEPH,DELPHI,L3,OPAL}. As an explicit example, we use the semileptonic channel to examine the effects from these backgrounds using {\tt whizard}~\cite{whizard}. We implement the simple cuts as follows~\cite{ILC:TGC2}: for the charged leptons $\ell = e,\, \mu,\, \tau$, $p_T > 10$ GeV and a separation of $> 5^\circ$ to the closet quark jets, $\slashed{E}_T > 10$ GeV, the visible mass $> 100$ GeV, the invariant masses of the decay products from $W$ bosons $< 105$ GeV, and the $W$ scattering angle $\cos\theta > -0.95$. We assume futher the $\tau$ leptons can be reconstructed with an efficiency of $80\%$. After these cuts, the total efficiency is about 91\%, as a result, the statistical errors in Table~\ref{tab:sensitivity} increase by about $11\% - 13\%$. For purely hadronic decays the background is significantly larger, while the leptonic channel is much cleaner. When these channels are considered, the efficiencies of paring of quark jets and assigning the $W$ charge or solving for the momenta of neutrinos has to be taken into account.

The precision of $W$ mass is expected to be 3 MeV at CEPC, and the beam energy uncertainty can reach up to 10 ppm, $\sim$1 MeV. The effects of radiative corrections and detector simulation are also well controlled, corresponding to a correction at roughly the same order~\cite{CEPC:slides,CEPC:preCDR}. We estimate roughly the effects due to these uncertainties, following the method in~\cite{ILC:TGC2}. For instance, we calculate the sensitivities of the anomalous couplings in presence of an energy variation of $\pm 2.4$ MeV, and compare them to the sensitivities in the standard scenario. The largest variation in the sensitivities is set as the corresponding systematic error due to the beam energy uncertainty. We find that the systematic corrections to the sensitivities are much smaller than the statistical errors (the former are expected to be of the order of magnitude of $\lesssim 10^{-5}$), and can be safely neglected. In other words, the statistical errors are expected to dominate the TGC measurements at CEPC. Detailed survey of the systematic errors needs a more dedicated and comprehensive study.

\subsection{contributions from different distributions}

To examine the contributions to the sensitivities from the distributions with respect to the five kinematic angles $\cos\theta$, $\cos\theta^\ast_{\ell}$, $\phi^\ast_{\ell}$, $\cos\theta^\ast_{\jmath}$, $\phi^\ast_{\jmath}$, we calculate
\begin{eqnarray}
\frac{\Delta\chi^2 (\Omega_k)}{\sum_k\Delta\chi^2 (\Omega_k)}
\end{eqnarray}
for the three $C$ and $P$ conserving aTGCs in Eq.~(\ref{eqn:tgc2operator}), and they are collected in Table~\ref{tab:contributions}. In this table, each of the entries stands for the corresponding proportion contributed to the total $\Delta \chi^2$ from one specific distribution, with each row summed to unity. We have omitted any potential correlations among the differential distributions which are expected to be small. The following qualitative features can be easily identified from the data in this table:
\begin{itemize}
  \item In all the three decay channels, the $\cos\theta$ distribution always dominate the sensitivities, consistent with implications of the magnitudes of the curves in Fig.~\ref{fig:omega} and the arguments in the previous section.
  \item For the semi-leptonic decays, distributions of the polar and azimuthal angles of the charged leptons are generally more important than the information from the hadron products. It is easily understood, as only the symmetric information from the jets can be used, as stated above.
  \item For the (semi-)leptonic decays, the TGCs are generally more sensitive to the azimuthal angles than to the polar angles, consistent with the magnitudes of the plots in Fig.~\ref{fig:omega}. An exception is the coupling $\Delta\kappa_\gamma$, for which the polar angles are also very important, cf. the middle panel in Fig.~\ref{fig:omega}.
  \item For the hadronic channel, the contribution from the decay information is small compared with the scattering angle $\cos\theta$, as expected.
  \item When all the channels are combined together, the contributions from different distributions are much similar to the semileptonic channel.
\end{itemize}
In short, the qualitative features of numerical estimations of the contributions of the five kinematic angles coincide with the theoretical arguments and predictions in the previous section. We stress that distributions of the decay products provide complementary information to the angle $\cos\theta$ and contribute sizably in constraining the aTGCs and relevant Dim-6 operators.

\begin{table}[t]
  \begin{center}
  \caption{Contributions to the sensitivities in Table~\ref{tab:sensitivity} of the three aTGCs in Eq.~(\ref{eqn:tgc2operator}) in the leptonic, semileptonic, hadronic and ``all'' channels from distributions of the five kinematic angles $\cos\theta$, $\cos\theta^\ast_\ell$, $\phi^\ast_\ell$, $\cos\theta^\ast_\jmath$ and $\phi^\ast_\jmath$. See text for details.}
  \label{tab:contributions}
  \begin{tabular}{ccccccc}
  \hline\hline
  \multicolumn{2}{c}{contributions} & $\cos\theta$ & $\cos\theta^\ast_\ell$ & $\phi^\ast_\ell$ & $\cos\theta^\ast_\jmath$ & $\phi^\ast_\jmath$ \\ \hline
  \multirow{3}{*}{leptonic} &
   $\Delta g_{1,Z}$         & 0.525 &  0.051 &  0.425 & - & - \\ \cline{2-7}
  &$\Delta \kappa_{\gamma}$ & 0.523 &  0.272 &  0.205 & - & - \\ \cline{2-7}
  &$\lambda_{\gamma}$       & 0.617 &  0.044 &  0.339 & - & - \\
  \hline
  \multirow{3}{*}{semi-leptonic} &
   $\Delta g_{1,Z}$         & 0.650 &  0.032 &  0.261 &  0.031 &  0.027 \\ \cline{2-7}
  &$\Delta \kappa_{\gamma}$ & 0.532 &  0.138 &  0.108 &  0.119 &  0.102 \\ \cline{2-7}
  &$\lambda_{\gamma}$       & 0.709 &  0.025 &  0.192 &  0.024 &  0.050 \\
  \hline
  \multirow{3}{*}{hadronic} &
   $\Delta g_{1,Z}$         & 0.850 & - & - &  0.080 &  0.070 \\ \cline{2-7}
  &$\Delta \kappa_{\gamma}$ & 0.546 & - & - &  0.244 &  0.210 \\ \cline{2-7}
  &$\lambda_{\gamma}$       & 0.827 & - & - &  0.056 &  0.118 \\
  \hline
  \multirow{3}{*}{all} &
   $\Delta g_{1,Z}$         & 0.722 &  0.020 &  0.167 &  0.048 &  0.042 \\ \cline{2-7}
  &$\Delta \kappa_{\gamma}$ & 0.538 &  0.081 &  0.065 &  0.170 &  0.147 \\ \cline{2-7}
  &$\lambda_{\gamma}$       & 0.755 &  0.015 &  0.117 &  0.036 &  0.076 \\
  \hline\hline
  \end{tabular}
  \end{center}
\end{table}

\section{Constraints at hadron colliders and future sensitivities}

The TGCs can also be probed directly at hadron colliders. So far, the Tevatron and the ATLAS and CMS collaborations have measured the charged anomalous couplings in the $WW$, $WZ$, and $W\gamma$ processes up to $\sqrt{s} = 8$ TeV~\cite{CDF,D0,D02,D03,ATLAS1,ATLAS2,ATLAS3,ATLAS4,ATLAS5,CMS1,CMS2,CMS3,CMS4}. There are also some speculation on the non-standard gauge couplings at the 14 TeV LHC~\cite{aTGC@LHC14a,aTGC@LHC14b}. These measurements are complementary to the EW precision tests, the accurate Higgs coupling probes, and all of these can be combined together to constrain the beyond SM physics~\cite{TGC:higgsdata,Ellis:2014jta}.

As a direct comparison, we consider simply the $WW$ production at the forthcoming LHC running at 14 TeV as an illustration. At parton level, the dominate channel is the process $q\bar{q} \rightarrow W^+W^-$, much like $e^+ e^- \rightarrow W^+W^-$ at lepton colliders, though the former obtains much larger radiative corrections~\cite{WWatLHCatNLO}. To suppress the huge QCD backgrounds, we focus on the purely leptonic decay channels $W \rightarrow e\nu, \, \mu\nu$. Due to the large missing energy carried away by the neutrinos and the unknown momenta carried by the initial quarks, we can not fully reconstruct the $W$ events. However, the observables in the transverse plane can yet be used to study the TGCs, e.g. the widely used leading $p_T$ of the charged lepton products~\cite{CDF,ATLAS1,ATLAS2,CMS3}. Analogous to the case at lepton colliders (cf. Tables~\ref{tab:contributions} and \ref{tab:contributions2}), at hadron colliders the azimuthal angles, or more specifically the difference between the azimuthal angles of charged leptons ($\Delta \phi_{ll}$) projected onto the transverse plane in the lab frame, are also very sensitive to the beyond SM TGCs and are very useful to help to constrain the couplings.

To explicitly demonstrate the arguments above and estimate the prospects for the constraints on the TGCs at LHC run II, we generate parton level events using {\tt MadGraph5}~\cite{MG5} at 14 TeV LHC and pass them to {\tt pythia}~\cite{pythia} and {\tt Delphes}~\cite{delphes}, for both the scenarios with and without beyond SM TGCs. Following~\cite{ATLAS2,Ellis:2014jta}, we implement the simple cuts below: for the charged leptons $l = e,\,\mu$, leading $p_T > 25$ GeV and subleading $p_T > 20$ GeV, $|\eta| < 2.5$, $\Delta R_{ll} > 0.4$, $m_{ll} > 15 (10)$ GeV, $\slashed{E}_T > 45 (15)$ GeV for the same (different) flavor channels, with the additional cut $\left| m_{ll} - M_Z \right| >15$ GeV for the same flavor channels.

\begin{figure}[t]
  \centering
  \includegraphics[width=0.45\textwidth]{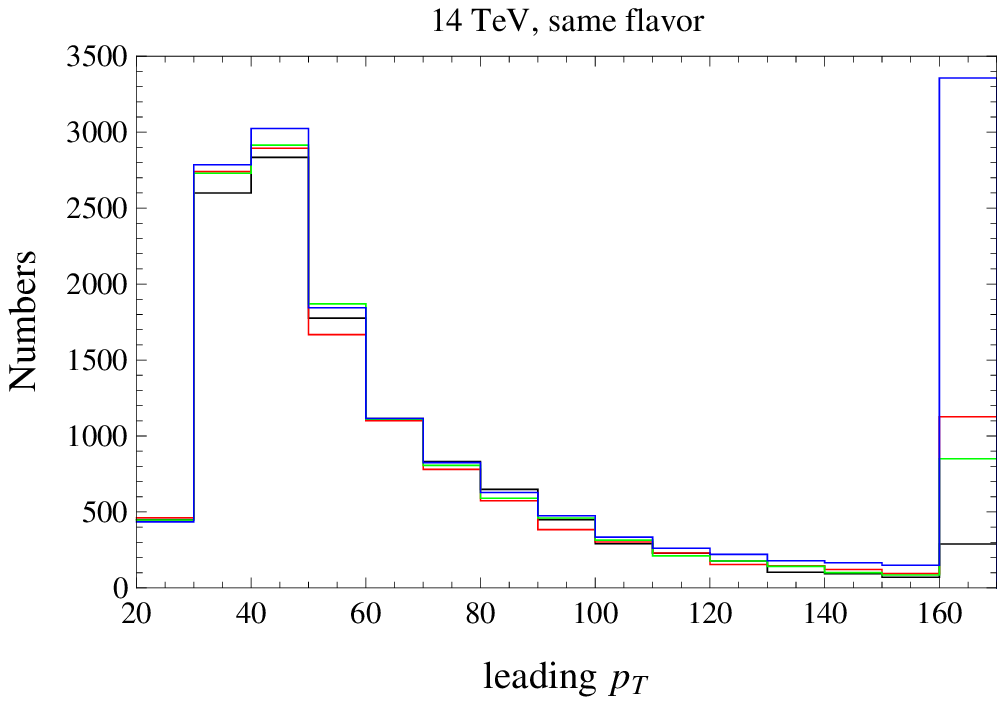}
  \includegraphics[width=0.45\textwidth]{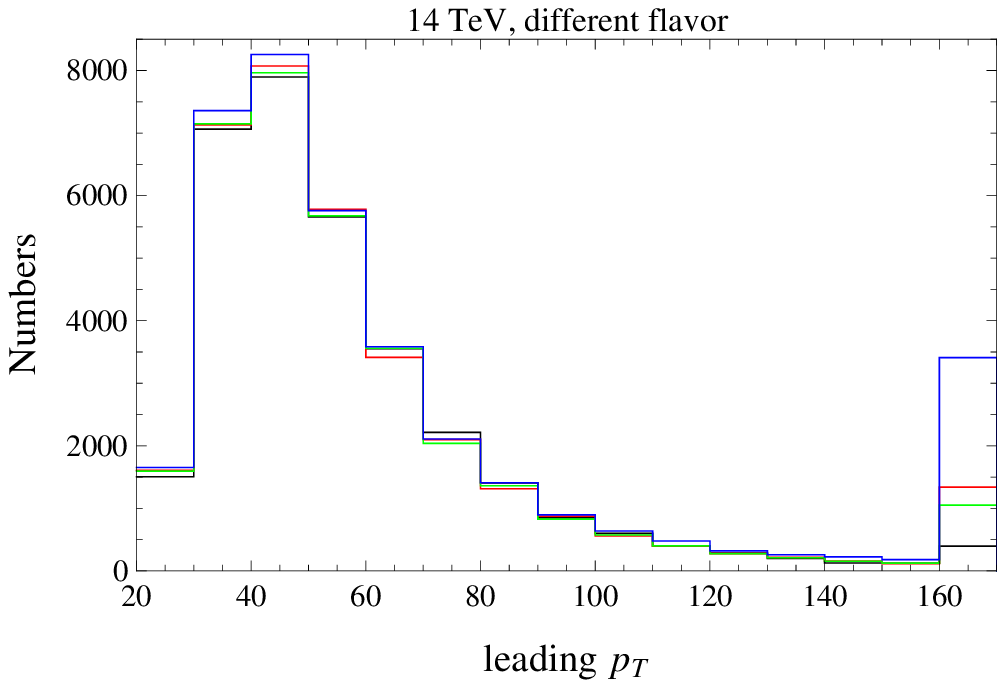} \\
  \includegraphics[width=0.45\textwidth]{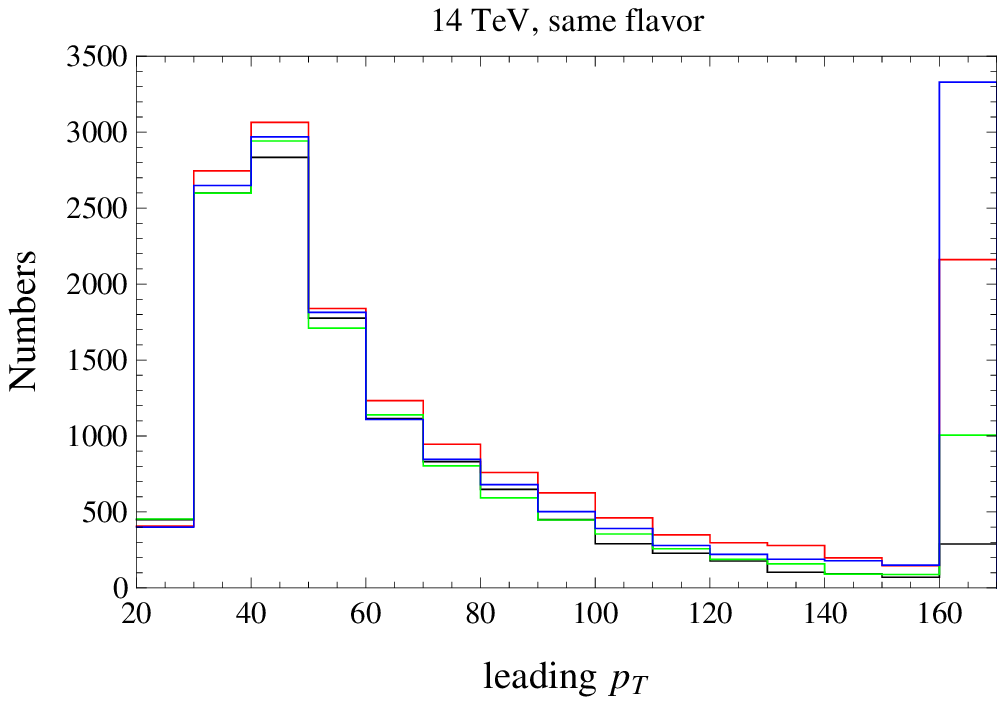}
  \includegraphics[width=0.45\textwidth]{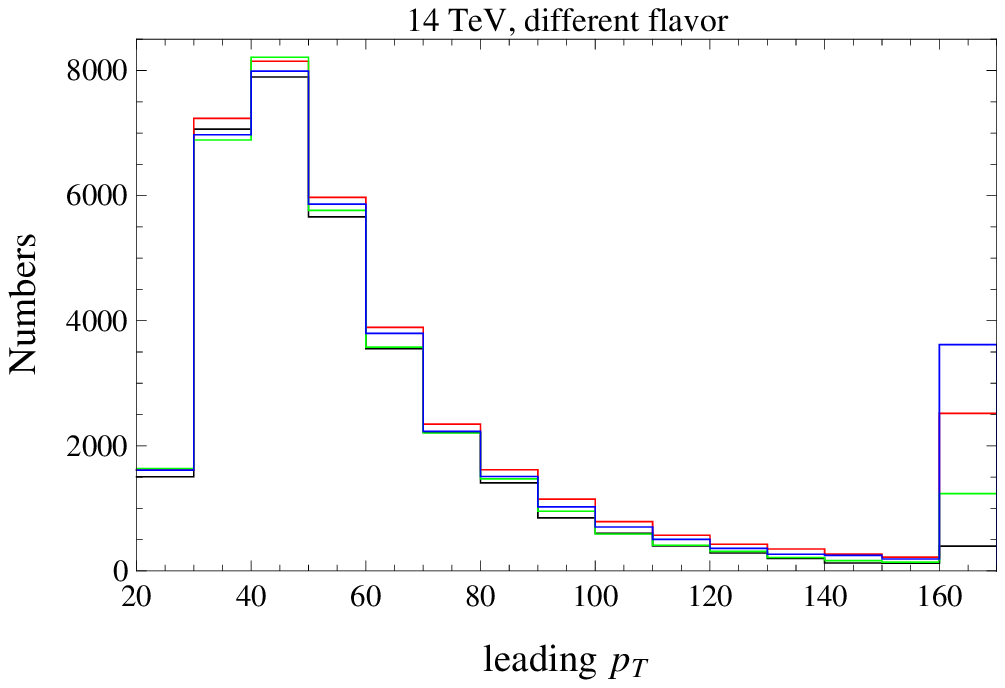}
  \vspace{-.5cm}
  \caption{Leading lepton $p_T$ distributions for purely leptonic decays of $pp \rightarrow WW$ at 14 TeV LHC. The two left panels are for the same flavor decays $ee$ and $\mu\mu$, while the two right for the different flavor channel $e\mu$. In the two upper pannels, the black, red, green and blue bins correspond respectively to the SM, $\Delta g_{1,\,Z} = 0.1$, $\Delta \kappa_\gamma = 0.2$ and $\lambda_\gamma = 0.1$, assuming EW relations between the aTGCs. In the two lower panels, the black, red, green and blue bins are respectively for the SM, $c_{HW} = 0.1$, $c_{HB} = 0.2$ and $c_{3W} = 0.1$. The last bin is an overflow bin.}
  \label{fig:pt}
\end{figure}

For illustration purpose, we choose some benchmark points beyond SM, i.e. $\Delta g_{1,\,Z} = 0.1$, $\Delta \kappa_\gamma = 0.2$ and $\lambda_\gamma = 0.1$ (assuming again the EW relations among the aTGCs) and $c_{HW} = 0.1$, $c_{HB} = 0.2$ and $c_{3W} = 0.1$. The leading $p_T$ distributions for all the seven scenarios above (the SM and the six beyond SM benchmark points) are presented in Fig.~\ref{fig:pt}, with the last bins are overflow bins. Obviously the anomalous couplings tend to generate large $p_T$ events. In presence of sizable non-standard couplings, the tails in these beyond SM scenarios are always much longer and fatter than in the SM, and the last bins are most sensitive to the non-standard TGCs.

\begin{figure}[t]
  \centering
  \includegraphics[width=0.45\textwidth]{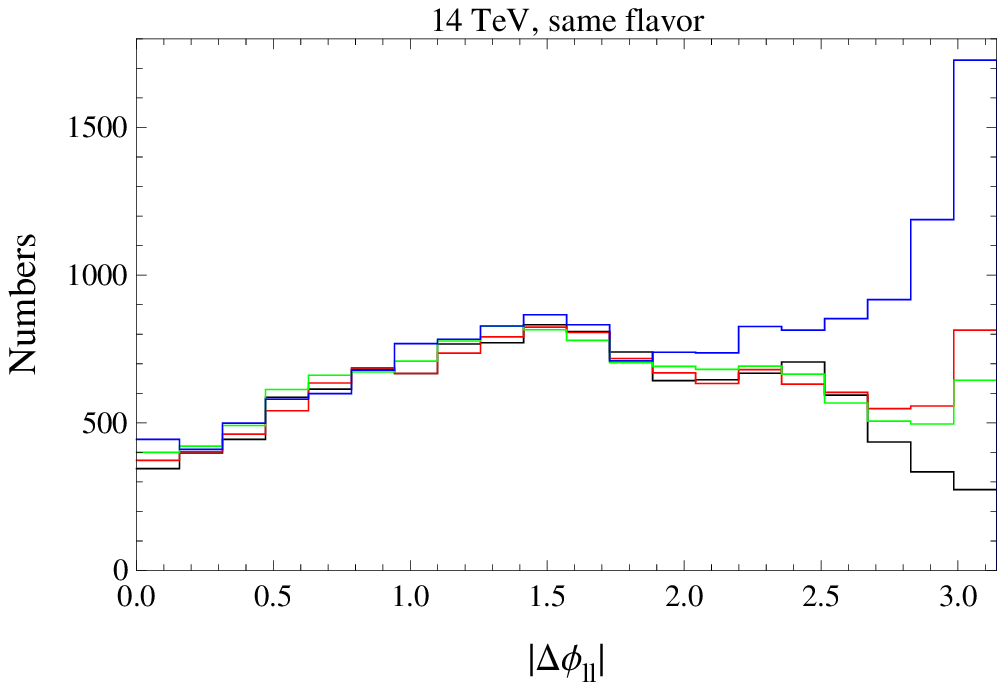}
  \includegraphics[width=0.45\textwidth]{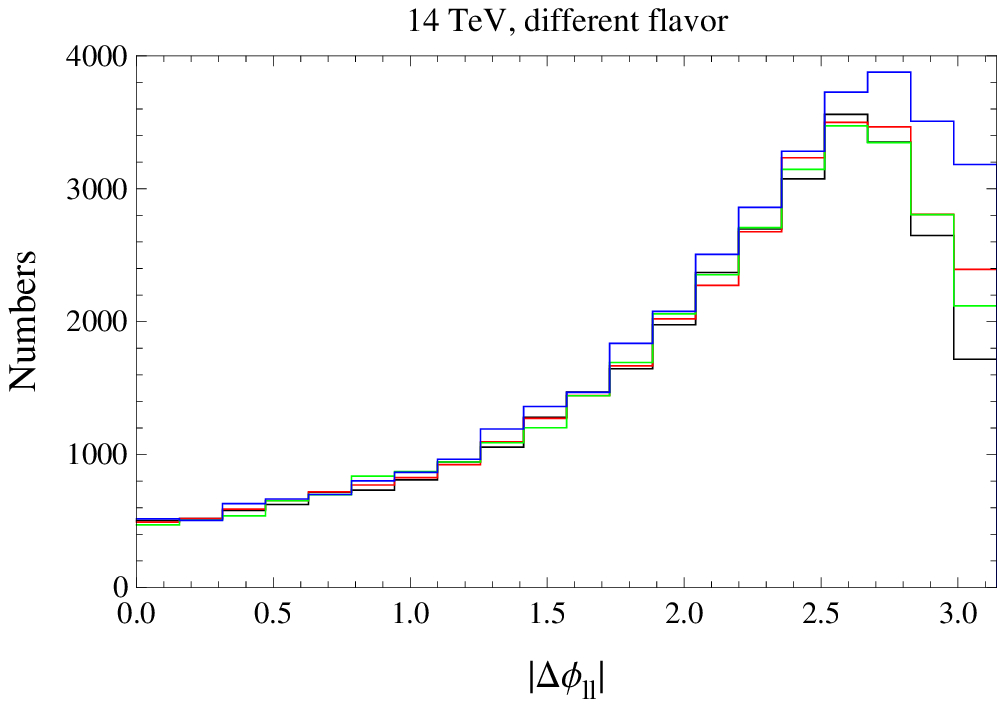} \\
  \includegraphics[width=0.45\textwidth]{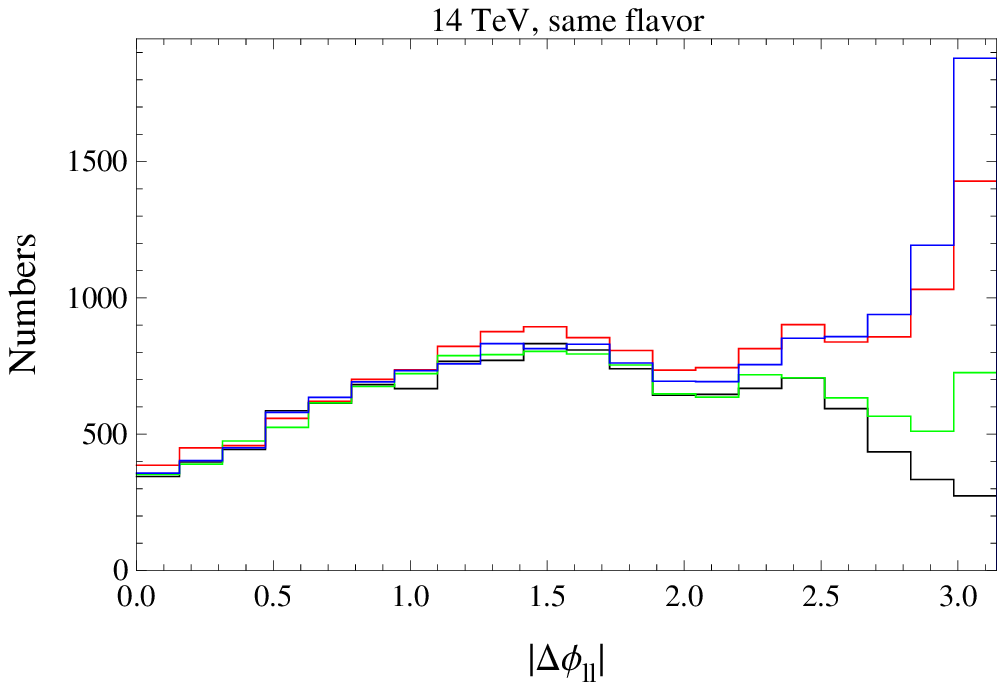}
  \includegraphics[width=0.45\textwidth]{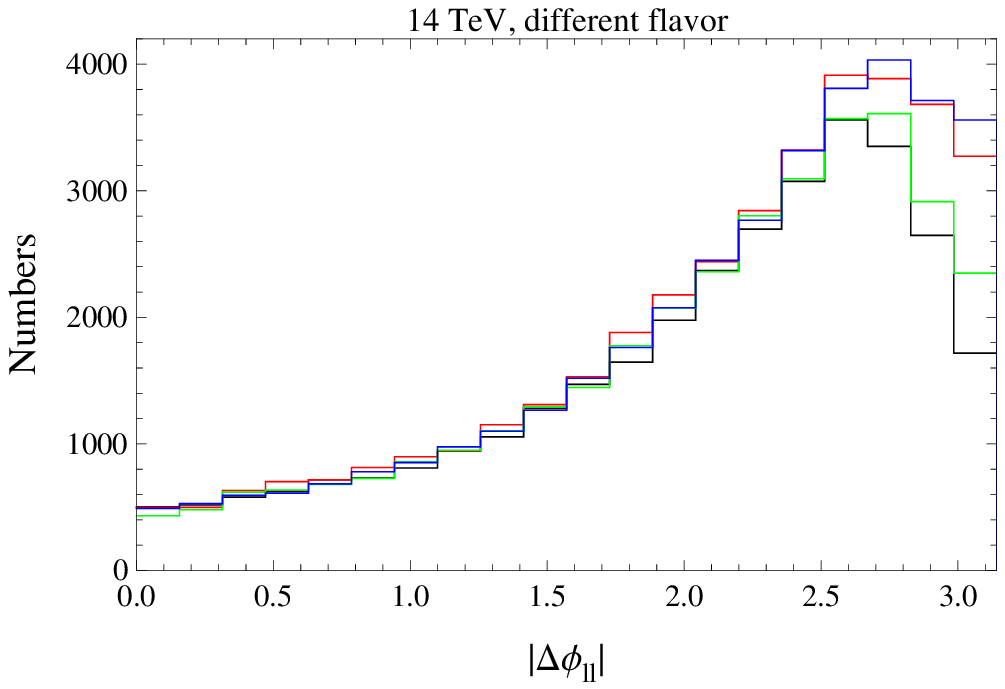}
  \vspace{-.5cm}
  \caption{The same as Fig.~\ref{fig:pt}, but for distributions of the azimuthal angle difference $\Delta \phi_{ll}$ for $WW$ events at 14 TeV LHC.}
  \label{fig:phi}
\end{figure}

We show in Fig.~\ref{fig:phi} the distributions of azimuthal angle difference $\Delta \phi_{ll}$ for the seven scenarios above. In presence of the beyond SM TGCs, the momenta of lepton products tend to be larger, thus we can expect more back-to-back like events, which explains qualitatively why the right few bins $\Delta \phi_{ll} \sim \pi$ of the plots in Fig.~\ref{fig:phi} are largely enhanced, especially for the same flavor decays. In light of the large excess of back-to-back events, combining the distributions of $\Delta \phi_{ll}$ with the leading $p_T$ could improve (moderately) the constraints on the anomalous couplings, though at hadron colliders the azimuthal angle $\Delta \phi_{ll}$ is strongly correlated to $p_T$. This is somewhat similar to the constraints at lepton colliders where the azimuthal angles of charged leptons are also very sensitive to the TGCs and contribute sizably to the constraints.

From the simulated events, we estimate roughly the constraints on the aTGCs and Dim-6 operators from the combined analysis of leading lepton $p_T$ and azimuthal angle $\Delta \phi_{ll}$ distributions. To take into account the large radiative corrections we use the the next-to-leading order total cross section of 124 pb for $pp \rightarrow W^+W^-$ at $\sqrt{s} = 14$ TeV to calculate the number of events~\cite{WWatLHCatNLO}. To optimise the constraints, in addition to the basic cuts above, we set the leading $p_T > 300$ GeV and $> 500$ GeV respectively for a luminosity of 300 fb$^{-1}$ and 3000 fb$^{-1}$ and further apply the cuts $\Delta \phi_{ll} > 170^\circ$, which improves moderately the constraints. To keep our events roughly below the cut-off scale $\tilde{\Lambda} \sim \tilde{g} M_W / \sqrt{c_i}$ in which the contributions from higher dimensional operators are suppressed \cite{Falkowski, Biekoetter:2014jwa}, it is important to set an upper limit for the leading $p_T$. In our cases, events beyond those upper limits are so rare unless one has a very weakly coupled theory $\tilde{g} < 0.3$\footnote{Notice here one can not use the unitarity bound for the $WW$ production to estimate the break down of EFT \cite{aTGC@LHC14b} since the weak couplings $q \bar{q} Z$ in the $WW$ production highly suppress the total rate, therefore the bound is essentially $4 \pi / g$ weaker than the real one.}. Nevertheless, one can lower the leading $p_T$ cut to make the EFT valid for models with even weaker coupling $\tilde{g}$, and the bounds on the dimension six operators would be moderately lower (for instance, bounds on $\mathcal{O}_{HW}$ by using $p_T > 160$ GeV would be 1.2 times smaller than those by using $p_T> 300$). We use conservatively only the same flavor decay products $ee$ and $\mu\mu$. Constraints on the anomalous couplings and Dim-6 operators are collected in Table~\ref{tab:LHC}. For a luminosity of 300 fb$^{-1}$, the limits are of the order of magnitude of $10^{-3}$. When the luminosity is ten times larger, the constraints go two or three times stronger. Notice that our analysis here are based on the simple cuts on the events with the same lepton flavors, there is actually a large improvement potential in a more elaborated study even in this di-lepton channel.

\begin{table}[t]
  \begin{center}
  \caption{$1\sigma$ constraints on the aTGCs and Dim-6 operators ($10^{-4}$) in Eq.~(\ref{eqn:tgc2operator}) from the same flavor leptonic decay channels of $pp \rightarrow W^+W^-$ at 14 TeV LHC with a luminosity of 300 fb$^{-1}$ and 3000 fb$^{-1}$. The numbers in parenthesis in the last three columns are the corresponding limits on the cut-off scale $\Lambda \sim M_W / \sqrt{c_i}$ in unit of TeV, cf Eq.~(\ref{Lagrangian}). See text for details.}
  \label{tab:LHC}
  \begin{tabular}{cccccccc}
  \hline\hline
   &   $\Delta g_{1,\,Z}$   & $\Delta \kappa_\gamma$ & $\lambda_\gamma$ & $\;\;$ & $c_{HW}$ & $c_{HB}$ & $c_{3W}$  \\ \hline
  300 fb$^{-1}$   & 23   & 73  &  17    && 14 (2.7)  & 73 (1.6)  & 17 (5.9) \\ \hline
  3000 fb$^{-1}$  & 11   & 30  & 5.7   && 6.3 (3.9)  & 30 (2.5)  & 5.7 (10) \\
  \hline\hline
  \end{tabular}
  \end{center}
\end{table}

\begin{figure}[t]
  \centering
  \includegraphics[width=0.6\textwidth]{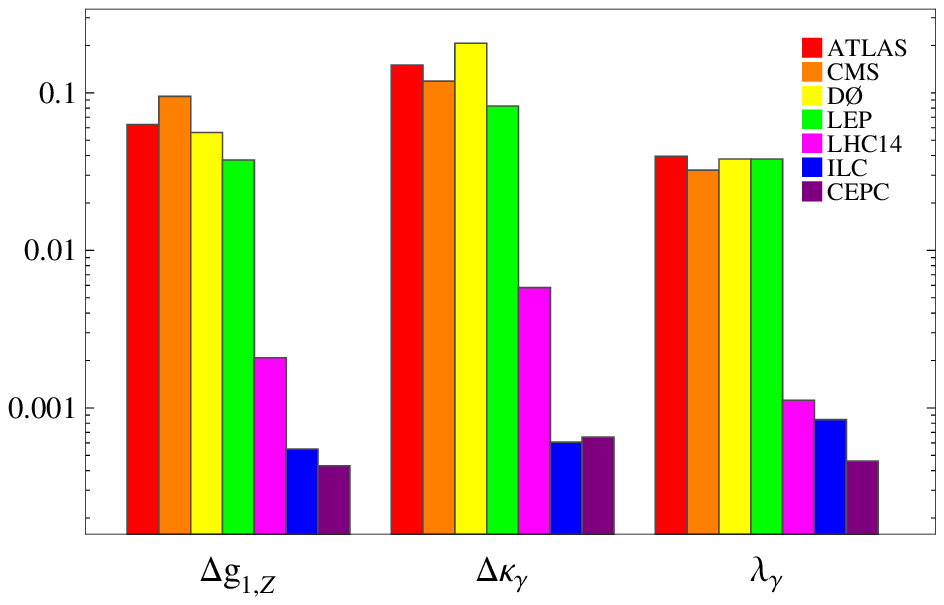}
  \vspace{-.5cm}
  \caption{Current and future 95\% confidence level constraints on the aTGCs. See text for details.}
  \label{fig:constraints}
\end{figure}

To close this section, we collect in Table~\ref{tab:comparison} and Fig.~\ref{fig:constraints} the current 95\% confidence level constraints on the aTGCs $\Delta g_{1,\,Z}$, $\Delta \kappa_\gamma$, $\lambda_\gamma$  from LEP, Tevatron and LHC and that from 14 TeV LHC and the future lepton colliders CEPC and ILC. The current lepton and hadron collider bounds are from~\cite{ATLAS5}\footnote{The current constraints on the five most general $C$ and $P$ violating aTGCs in Eq.~(\ref{eqn:aTGC5}) from LEP, Tevatron and LHC can be found in Ref.~\cite{aTGC:comparison1,aTGC:comparison2}.}, for which we do not collect all the limits given in this reference but pick out the most stringent ones for each of experimental groups. In Fig.~\ref{fig:constraints}, the data for LHC 14 TeV assume a luminosity of 3000 fb$^{-1}$, the limits for CEPC use only the semileptonic channel, for ILC they are the limits at $\sqrt{s} = 500$ GeV with a luminosity of 500 pb$^{-1}$ from~\cite{ILC:TDR}.\footnote{The ILC constraints are obtained by using the $\cos\theta_W$ distribution and the single particle SDM in the semileptonic channel~\cite{ILC:TGC1,ILC:TGC2} while in this work the differential cross sections with respect to the five kinematic angles are used to set the limits on CEPC.} At future lepton colliders, using more decay channels, higher energies and larger luminosity can improve further the constraints in this figure. Comparing na\"ively the limits in this table, the 14 TeV LHC and future lepton colliders can improve the limits on the aTGCs by one to two orders of magnitude. Benefitting from the huge integrated luminosity, CEPC can get similar constraints on the TGCs in comparison with that from ILC or can do even better. Accumulating a larger amount of data, which is designed up to 10 ab$^{-1}$ at $\sqrt{s} = 240$ GeV, it is expected that the TLEP (recently renamed as FCC-ee) can improve further constraints on the aTGCs~\cite{Gomez-Ceballos:2013zzn}. To get rough constraints on the relevant Dim-6 operators, it is a viable to follow the arguments in section~\ref{sec:distribution} and use simply the matrix $V_{ij}$ to convert the limits on the aTGCs to that on the Dim-6 coefficients.

\begin{table}[t]
  \begin{center}
  \caption{95\% confidence level constraints on the aTGCs from the current and future colliders. For ATLAS the constraints on $\Delta g_{1,\,Z}$ and $\lambda_\gamma$ are from~\cite{ATLAS5} and the limit on $\Delta \kappa_\gamma$ is from~\cite{ATLAS2}. The CMS constraint on $\Delta g_{1,\,Z}$ is from Ref.~\cite{CMS3} and the other two CMS limits are from~\cite{CMS1}. The LHC 14 TeV constraints assume a integrated luminosity of 3000 (300) fb$^{-1}$. For ILC they are the limits at the center-of-mass energy of 500 (800) GeV with a luminosity of 500 (1000) fb$^{-1}$, while for CEPC we list the constraints coming from the semileptonic channel (combining all the available channels), cf. Table~\ref{tab:sensitivity}. }
  \label{tab:comparison}
  \begin{tabular}{lccc}
  \hline\hline
    & $\Delta g_{1,\,Z}$ & $\Delta \kappa_\gamma$ & $\lambda_\gamma$ \\ \hline
  ATLAS~\cite{ATLAS2,ATLAS5} & $[-0.055, 0.071]$  & $[-0.150, 0.150]$  & $[-0.039, 0.040]$  \\ \hline
  CMS~\cite{CMS1,CMS3}       & $[-0.095, 0.095]$  & $[-0.104, 0.134]$  & $[-0.036, 0.028]$  \\ \hline
  D0~\cite{D03}              & $[-0.031, 0.081]$  & $[-0.158, 0.255]$  & $[-0.034, 0.042]$  \\ \hline
  LEP~\cite{LEP}             & $[-0.054, 0.021]$  & $[-0.099, 0.066]$  & $[-0.059, 0.017]$  \\ \hline
  LHC14 & $\begin{matrix} [-0.0021,0.0021] \\ ([-0.0045,0.0045]) \end{matrix}$ &
          $\begin{matrix} [-0.0058,0.0058] \\ ([-0.014,0.014])   \end{matrix}$ &
          $\begin{matrix} [-0.0011,0.0011] \\ ([-0.0033,0.0033]) \end{matrix}$ \\ \hline
  ILC~\cite{ILC:TDR} & $\begin{matrix} [-0.00055,0.00055] \\ ([-0.00035,0.00035]) \end{matrix}$ &
                       $\begin{matrix} [-0.00061,0.00061] \\ ([-0.00037,0.00037]) \end{matrix}$ &
                       $\begin{matrix} [-0.00084,0.00084] \\ ([-0.00051,0.00051]) \end{matrix}$ \\ \hline
  CEPC  & $\begin{matrix} [-0.00043,0.00043] \\ ([-0.00031,0.00031]) \end{matrix}$ &
          $\begin{matrix} [-0.00065,0.00065] \\ ([-0.00045,0.00045]) \end{matrix}$ &
          $\begin{matrix} [-0.00046,0.00046] \\ ([-0.00033,0.00033]) \end{matrix}$ \\
  \hline\hline
  \end{tabular}
  \end{center}
\end{table}

\section{Complementarity on the measurements of the TGCs, EW precision observable and the Higgs couplings}

As stated in the introduction, besides the direct measurements in the di-boson channels at lepton and hadron colliders, the TGCs can also be indirectly probed by the EW precision data and the Higgs data. The oblique parameter, TGCs and the Higgs couplings can be simply transmitted to each other by redefining the gauge fields, or from integration by parts and equation of motion~\cite{Grojean}:
\begin{eqnarray}
- \frac{2g s c v^2}{\alpha} \mathcal{O}_S - \frac{g' v^2}{\alpha} \mathcal{O}_T + g'  \mathcal{O}_{hf}^Y
&=& 2 g' \mathcal{O}_{HB}  -  g' \mathcal{O}_{h2} + \frac{g'}{2}  \mathcal{O}_{BB} - \frac{g'}{2} \mathcal{O}_{h3}, \nonumber \\
- \frac{4 g' s c v^2}{\alpha} \mathcal{O}_S +g (\mathcal{O}_{hl}^t+\mathcal{O}_{hq}^t)
&=&  4 g \mathcal{O}_{HW} - 6 g \mathcal{O}_{h2}  +g \mathcal{O}_{WW} - g \mathcal{O}_{h3},
\end{eqnarray}
where $\mathcal{O}_{HW}$, $\mathcal{O}_{HB}$ are defined in Eq.~(\ref{Lagrangian}) and
\begin{eqnarray}
\mathcal{O}_{S} &\equiv& (\alpha/4 s  c v^2) \mathcal{O}_{WB} = (\alpha/4 s  c v^2)(H^\dagger \sigma^a H) W_{\mu \nu}^a B^{\mu \nu} \,, \\
\mathcal{O}_T &\equiv& -(2 \alpha/v^2) \mathcal{O}_h = -(2 \alpha/v^2) |H^\dagger D_\mu H|^2 \,,
\end{eqnarray}
so that their coefficients are the $S$, $T$ parameter.
\begin{eqnarray}
\mathcal{O}_{hf}^s &\equiv& i (H^\dagger D_\mu H)(\bar{f}\gamma^\mu f)+h.c. \\
\mathcal{O}^t_{hf} &\equiv& i (H^\dagger \sigma^a D_\mu H)(\bar{f}\gamma^\mu \sigma^a f)+h.c. \\
\mathcal{O}_{BB}&\equiv& H^\dagger H \left(B_{\mu\nu}\right)^2 \\
\mathcal{O}_{h2} &\equiv& H^\dagger H D^\mu H^\dagger D_\mu H \\
\mathcal{O}_{h3} &\equiv& H^\dagger H \left(H^\dagger D^2 H + (D^2 H^\dagger)H\right) \\
\mathcal{O}_{WW} &\equiv& H^\dagger H \left(W^a_{\mu\nu}\right)^2 \,.
\end{eqnarray}
For the Higgs measurement sensitivity, we consider the $hZZ$ coupling constraints from $\mathcal{O}_{h2}$ since its contribution has the same form as the SM one (no derivative couplings) and we can obtain from the relations between different sensitivities that
\begin{eqnarray}
 \nonumber
c_{HB} &\sim&  \frac{\alpha g^2}{4 c^2} \Delta S \sim  \frac{\alpha g^2}{2} \Delta T  \sim {2} c_{h2} \sim {g^2} \Delta g_{hZZ} / g_{hZZ},
         \label{eq:l3} \\
c_{HW} &\sim& \frac{\alpha g^2}{4 s^2} \Delta S  \sim \frac{2}{3} c_{h2} \sim \frac{g^2}{3} \Delta g_{hZZ} / g_{hZZ},
         \label{eq:l4}
\end{eqnarray}

From Ref.~\cite{LTWang}, one can obtain the EW and Higgs precision $\Delta g_{hZZ} / g_{hZZ}$ for high luminosity LHC (HL-LHC), CEPC, ILC and TLEP. Therefore, we can easily compare the direct constraints on TGCs with EW and Higgs precision $\Delta g_{hZZ} / g_{hZZ}$ by re-shifting them into TGCs from Eq. (\ref{eq:l4}). 
The results are listed in Table~\ref{tab:complementarity}. From this very rough examinations, it is interesting to see that the sensitivities of TGCs are comparable with the EW precision for certain new physics operators like $\mathcal{O}_{HW}$, which strikingly raises the importance of improving the TGC measurements in the future. More detailed and deeper study of this complementarity on the distinct measurements of the TGCs and the global new physics fits at LHC and future colliders will be performed in a separate paper. Notice that the FCC-ee also has a run option at a center-of-mass energy of $\sim 250$ GeV, similar to the CEPC one, with possibly larger integrated luminosity, hence it is expected that the sensitivity on TGCs would be improved by a certain amount.

\begin{table}[t]
  \begin{center}
  \caption{Comparison of the direct and indirect measurements of the TGCs from future data. The limits from high luminosity LHC at 14 TeV assume a luminosity of 3000 fb$^{-1}$, and the CEPC data use only the semileptonic channel, cf. Tables~\ref{tab:sensitivity} and \ref{tab:LHC}. The prospects for the oblique parameters $S$, $T$ and the $hZZ$ coupling uncertainties at HL-LHC and CEPC are from~\cite{LTWang}, and are translated to constraints on the corresponding Dim-6 operators.}
  \label{tab:complementarity}
  \begin{tabular}{lccc}
  \hline\hline
  & future prospects & $c_{HW}$ & $c_{HB}$ \\ \hline
  HL-LHC & - & $6.3\times 10^{-4}$ & $3\times 10^{-3}$ \\ \hline
  CEPC  & - & $1.2\times 10^{-4}$ & $3.3\times 10^{-4}$ \\ \hline
  $S$: HL-LHC & 0.13 & $5\times 10^{-4}$ & $1.4\times10^{-4}$ \\ \hline
  $T$: HL-LHC & 0.09 & $-$ & $1.6\times 10^{-4}$ \\ \hline
  $\frac{\Delta g_{hZZ}}{g_{hZZ}}$: HL-LHC & 0.03 & $4.5\times 10^{-3}$ & $1.3\times 10^{-2}$ \\ \hline
  $S$: CEPC & 0.04 & $1.6\times 10^{-4}$ & $4.2\times10^{-5}$ \\ \hline
  $T$: CEPC & 0.03 & $-$ & $5.3\times 10^{-5}$ \\ \hline
  $\frac{\Delta g_{hZZ}}{g_{hZZ}}$: CEPC & 0.002 & $3\times 10^{-4}$ & $9\times 10^{-4}$ \\
  \hline\hline
  \end{tabular}
  \end{center}
\end{table}

\section{Conclusion}

In the era of precision measurements of the SM couplings among the scalars, fermions and gauge boson, the triple couplings among the SM EW gauge bosons is an essential part to test the SM in the gauge sector and set constraints on precision electroweak and Higgs physics, which can give us a powerful guidance on searching for new physics beyond the SM. 

$WW$ process is the primary channel at lepton colliders to measure directly the charged triple couplings. Kinematically this process can be described by five angles, including those for the decay products of $W$ boson. In this work we use the five angles to study the lepton collider constraints on the anomalous triple gauge couplings and the relevant three dimension-6 operators $c_{HW}$, $c_{HB}$ and $c_{3W}$ in the $C$ and $P$ conserving sector. We obtain numerically and graphically the effects of anomalous triple couplings on the differential cross sections with respect to the five kinematic angles. From the plots in Fig.~\ref{fig:omega} one can identify qualitatively which distributions are more sensitive to the anomalous couplings and which coupling is expected to be most severely constrained.

We calculate systematically the statistical errors of the anomalous couplings and dimension-6 operators at CEPC with $\sqrt{s} = 240$ GeV and a luminosity of 5 ab$^{-1}$, using simply the shapes of the differential cross sections with respect to the five angles, in all the leptonic, semileptonic and hadronic decay channels of $W$ pairs. The sensitivities, collected in Tables~\ref{tab:sensitivity} and \ref{tab:sensitivity2}, can reach up to the order of magnitude of $10^{-4}$, comparable to that on ILC or even better for some couplings. We find the information from the decay products are complementary to the $W$ scattering angle $\cos\theta$ and contribute sizably to the sensitivities. The importance of the decay information and the corresponding contribution depend largely on the decay channels and the anomalous couplings involved, which are collected in Tables~\ref{tab:contributions} and \ref{tab:contributions2}.

We have also investigated the constraints at hadron colliders, the 14 TeV LHC and estimate roughly the sensitivities in the di-lepton channels for $WW$ production at the 14 TeV LHC with a luminosity of 300 fb$^{-1}$ and 3000 fb$^{-1}$. Depending on the luminosity and the anomalous couplings involved, the constraints are at the level of $10^{-2}$ to $10^{-3}$, which are collected in Tables~\ref{tab:LHC} and \ref{tab:LHC2}. The constraints are mainly due to the excess of events for high leading lepton $p_T$ in presence of the non-standard couplings. The azimuthal angle difference $\Delta \phi_{ll}$ of the charged leptons also contribute moderately to the constraints. At the end, we collect the current and future constraints on the anomalous triple gauge couplings from both future lepton colliders and the 14 TeV LHC, and compare their sensitivities with the precision EW and Higgs couplings  in terms of dimension six operators. It is promising that constraints on the charged triple gauge couplings can be improved by two orders of magnitude and reach the order of magnitude of $10^{-4}$. The sensitivity gap between electroweak precision and triple gauge boson precision can be significantly decreased to less than one order of magnitude or even less at the 14 TeV HL-LHC and both the two sensitivities can be improved at future lepton colliders such as CEPC, which allows us to reconsider those triple gauge boson constrains on the EW precision physics in the future. Finally it is worth mentioning that the future FCC-ee data at 240 GeV with a potentially larger integrated luminosity than CEPC, as well as the TLEP-$W$ and TLEP-$Z$ data, could improve further the constraints on charged triple gauge couplings and other new interactions beyond the standard model.

\section*{Acknowledgements}
We would like to thank Huaqiao Zhang for helpful discussions and reading the manuscript.

\bigskip
\noindent{\it Note added.} \ While this paper is being finalized, Ref.~\cite{Wells:2015eba} appeared which also discussed $e^+ e^- \rightarrow W^+ W^-$ and has some overlap with our paper. Nevertheless, we also consider the constraints from 14 TeV LHC and compare the aTGCs constraints with those from EW precision observables and Higgs couplings.

\appendix
\section{Constraints on the five most general $C$ and $P$ conserving aTGCs}
For completeness we collect here the main results in this work for the five most general $C$ and $P$ conserving aTGCs $\Delta g_{1,Z}$, $\Delta \kappa_\gamma$, $\Delta\kappa_Z$, $\lambda_\gamma$ and $\lambda_Z$. The five $b_i$ at $\sqrt{s} = 240$ GeV are, respectively,
\begin{alignat}{2}
\label{eqn:bi2}
b_1 &= b (\Delta g_{1,Z})         &&= -0.0387 \,, \nonumber \\
b_2 &= b (\Delta \kappa_{\gamma}) &&= -0.178 \,, \nonumber \\
b_3 &= b (\Delta \kappa_{Z})      &&= -0.0813 \,, \nonumber \\
b_4 &= b (\lambda_{\gamma})       &&= -0.0884 \,, \nonumber \\
b_5 &= b (\lambda_{Z})            &&= -0.0154 \,.
\end{alignat}
The analytical and numerical expressions for the corresponding $\omega_i (\Omega_k)$ are collected in Appendices~\ref{sec:abi} and \ref{sec:bi}. The one-parameter constraints at CEPC are listed in Table~\ref{tab:sensitivity2}. Comparing the data in Table~\ref{tab:sensitivity} and \ref{tab:sensitivity2}, we find that: (i) The relation $\lambda_\gamma = \lambda_Z$ combines the constraints on the two couplings together, consequently the limits on $\lambda_\gamma = \lambda_Z$ in Table~\ref{tab:sensitivity} are more stringent than the two parameters separately in Table~\ref{tab:sensitivity2}. (ii) Likewise, due to the relation $\Delta \kappa_Z = \Delta g_{1,Z} -  \tan^2\theta_W \Delta \kappa_\gamma$, $\Delta g_{1,\,Z}$ absorbs some sensitivities on $\Delta \kappa_Z$ and is more severely constrained in Table~\ref{tab:sensitivity}. (iii) On the contrary, $\Delta \kappa_\gamma$ is anti-correlated to $\Delta \kappa_Z$, thus it is less stringent constrained in Table~\ref{tab:sensitivity}. The correlation matrix between the five aTGCs $\Delta g_{1,\,Z}$, $\Delta \kappa_{\gamma,\,Z}$ and $\lambda_{\gamma,\,Z}$ are
\begin{eqnarray}
\label{eqn:rho-all}
\rho^{} = \left( \begin{matrix}
  1 &  &&& \\
  0.827 &  1 &  && \\
  0.966 &  0.918 &  1 &  & \\
  0.905 &  0.954 &  0.940 &  1 &   \\
  0.948 &  0.696 &  0.895 &  0.839 &  1
\end{matrix} \right) \,,
\end{eqnarray}
where all the three channels are combined together. The analogue of Table~\ref{tab:contributions} and \ref{tab:LHC} for the five aTGCs are respectively given in Table~\ref{tab:contributions2} and \ref{tab:LHC2}.

\begin{table}[t]
  \begin{center}
  \caption{The same as Table~\ref{tab:sensitivity}, but for the five aTGCs in Eq.~(\ref{eqn:aTGC5}).}
  \label{tab:sensitivity2}
  \begin{tabular}{cccccc}
  \hline\hline
  channels & $\Delta g_{1,Z}$ & $\Delta \kappa_\gamma$ & $\Delta\kappa_Z$ & $\lambda_\gamma$ & $\lambda_Z$ \\ \hline
  leptonic     & 14.49 & 8.02 & 9.82 & 12.70 & 12.00 \\ \hline
  semileptonic &  5.54 & 2.71 & 3.59 &  4.32 &  4.63 \\ \hline
  hadronic     &  6.56 & 2.74 & 4.00 &  4.40 &  5.65 \\ \hline
  all          &  4.06 & 1.87 & 2.58 &  3.00 &  3.44 \\
  \hline\hline
  \end{tabular}
  \end{center}
\end{table}

\begin{table}[t]
  \begin{center}
  \caption{The same as Table~\ref{tab:contributions}, but for the five aTGCs in Eq.~(\ref{eqn:aTGC5}). }
  \label{tab:contributions2}
  \begin{tabular}{ccccccc}
  \hline\hline
  \multicolumn{2}{c}{contributions} & $\cos\theta$ & $\cos\theta^\ast_\ell$ & $\phi^\ast_\ell$ & $\cos\theta^\ast_\jmath$ & $\phi^\ast_\jmath$ \\ \hline
  \multirow{5}{*}{leptonic} &
   $\Delta g_{1,Z}$         & 0.483 &  0.032 &  0.484 & - & - \\ \cline{2-7}
  &$\Delta \kappa_{\gamma}$ & 0.572 &  0.233 &  0.194 & - & - \\ \cline{2-7}
  &$\Delta \kappa_{Z}$      & 0.552 &  0.068 &  0.381 & - & - \\ \cline{2-7}
  &$\lambda_{\gamma}$       & 0.675 &  0.130 &  0.196 & - & - \\ \cline{2-7}
  &$\lambda_{Z}$            & 0.449 &  0.008 &  0.542 & - & - \\
  \hline\hline
  \multirow{5}{*}{semi-leptonic} &
   $\Delta g_{1,Z}$         & 0.629 &  0.020 &  0.312 &  0.020 &  0.020 \\ \cline{2-7}
  &$\Delta \kappa_{\gamma}$ & 0.586 &  0.119 &  0.102 &  0.104 &  0.089 \\ \cline{2-7}
  &$\Delta \kappa_{Z}$      & 0.662 &  0.040 &  0.226 &  0.038 &  0.033 \\ \cline{2-7}
  &$\lambda_{\gamma}$       & 0.699 &  0.067 &  0.102 &  0.057 &  0.075 \\ \cline{2-7}
  &$\lambda_{Z}$            & 0.599 &  0.006 &  0.358 &  0.003 &  0.034 \\
  \hline\hline
  \multirow{5}{*}{hadronic} &
   $\Delta g_{1,Z}$         & 0.889 & - & - &  0.055 &  0.056 \\ \cline{2-7}
  &$\Delta \kappa_{\gamma}$ & 0.602 & - & - &  0.214 &  0.184 \\ \cline{2-7}
  &$\Delta \kappa_{Z}$      & 0.823 & - & - &  0.095 &  0.081 \\ \cline{2-7}
  &$\lambda_{\gamma}$       & 0.725 & - & - &  0.119 &  0.155 \\ \cline{2-7}
  &$\lambda_{Z}$            & 0.890 & - & - &  0.008 &  0.102\\
  \hline\hline
  \multirow{5}{*}{all} &
   $\Delta g_{1,Z}$         & 0.718 &  0.013 &  0.205 &  0.032 &  0.032 \\ \cline{2-7}
  &$\Delta \kappa_{\gamma}$ & 0.592 &  0.070 &  0.061 &  0.149 &  0.128 \\ \cline{2-7}
  &$\Delta \kappa_{Z}$      & 0.722 &  0.025 &  0.142 &  0.059 &  0.050 \\ \cline{2-7}
  &$\lambda_{\gamma}$       & 0.710 &  0.039 &  0.060 &  0.083 &  0.108 \\ \cline{2-7}
  &$\lambda_{Z}$            & 0.696 &  0.004 &  0.239 &  0.004 &  0.057 \\
  \hline\hline
  \end{tabular}
  \end{center}
\end{table}

\begin{table}[t]
  \begin{center}
  \caption{The same as Table~\ref{tab:LHC}, but for the five most general $C$ and $P$ conserving aTGCs ($10^{-4}$) in Eq.~(\ref{eqn:aTGC5}) from $pp \rightarrow W^+W^-$ at 14 TeV LHC. }
  \label{tab:LHC2}
  \begin{tabular}{cccccc}
  \hline\hline
   &   $\Delta g_{1,\,Z}$   & $\Delta \kappa_\gamma$ & $\Delta \kappa_Z$ & $\lambda_\gamma$ & $\lambda_Z$   \\ \hline
  300 fb$^{-1}$   & 318   & 50  &  20    & 49 & 19 \\ \hline
  3000 fb$^{-1}$  & 180   & 21  & 9.7   & 24 & 9.0  \\
  \hline\hline
  \end{tabular}
  \end{center}
\end{table}

\section{Analytical expressions for the coefficients $\omega_i$}
\label{sec:abi}

All the differential cross sections with respect to the five production and decay angles can be obtained from Eq.~(\ref{eqn:diff}) by integrating out some of the angles. The SM differential cross sections read
\begin{eqnarray}
\frac{{\rm d}\sigma_0}{{\rm d}\cos\theta }
&=& \frac{\beta}{32\pi s}
\sum_{\lambda\tau_1\tau_2} \left[ F^{(\lambda)}_{\tau_1\tau_2} F^{(\lambda)\ast}_{\tau_1\tau_2} \right]_0 \,, \\
\frac{{\rm d}\sigma_0 }
{ {\rm d}\cos\theta^\ast_{1,\,2} }
&=& {\rm BR} \cdot \frac{\beta}{32\pi s} \left( \frac{3}{4} \right)^2
\sum_{\lambda\tau_1\tau_2}
\int {\rm d} \cos\theta \int {\rm d} \cos\theta^\ast_{2,\,1} \,
\left[ F^{(\lambda)}_{\tau_1\tau_2} F^{(\lambda)\ast}_{\tau_1\tau_2} \right]_0 \nonumber \\
\label{eqn:b2}
&& \times D_{\tau_1 \tau_1} (\theta^\ast_1) D_{\tau_2 \tau_2} (\pi-\theta^\ast_2) \,, \\
\frac{{\rm d}\sigma_0 }{ {\rm d}\phi^\ast_{1} }
&=& {\rm BR} \cdot \frac{\beta}{32\pi s} \left( \frac{3}{8\pi} \right)
\sum_{\lambda\tau_1\tau'_1\tau_2}
\int {\rm d} \cos\theta \int {\rm d} \cos\theta^\ast_{1} \,
\left[ F^{(\lambda)}_{\tau_1\tau_2} F^{(\lambda)\ast}_{\tau'_1\tau_2} \right]_0 \nonumber \\
&& \times D_{\tau_1 \tau'_1} (\theta^\ast_1, \phi^\ast_1)  \,, \\
\frac{{\rm d}\sigma_0 }{ {\rm d}\phi^\ast_{2} }
&=& {\rm BR} \cdot \frac{\beta}{32\pi s} \left( \frac{3}{8\pi} \right)
\sum_{\lambda\tau_1\tau_2\tau'_2}
\int {\rm d} \cos\theta \int {\rm d} \cos\theta^\ast_{2} \,
\left[ F^{(\lambda)}_{\tau_1\tau_2} F^{(\lambda)\ast}_{\tau_1\tau'_2} \right]_0 \nonumber \\
&& \times D_{\tau_2 \tau'_2} (\pi-\theta^\ast_2, \pi+\phi^\ast_2)  \,,
\end{eqnarray}
where the scattering amplitudes $F^{(\lambda)}_{\tau_1\tau_2}$ are generally linear functions of the anomalous couplings $\alpha_i = \Delta g_{1,\,V},\, \Delta\kappa_V,\, \lambda_V,\, g_5^V,\, g_4^V,\, \tilde{\kappa}_V,\, \tilde{\lambda}_V$ in Eq.~(\ref{eqn:lagrangianTGC}), $[\cdots]_0$ means that the we take only the SM contribution with $\alpha_i \rightarrow 0$. In Eq.~(\ref{eqn:b2}), dependence of the decay matrix $D$ on the azimuthal angles has been integrated out
\begin{eqnarray}
D (\theta) = \left( \begin{matrix}
\frac12 (1-\cos\theta)^2 && \\
& \sin^2\theta & \\
&& \frac12 (1+\cos\theta)^2
\end{matrix} \right) \,.
\end{eqnarray}

According to Refs.~\cite{Bilenky2,Gounaris}, the amplitudes $F^{(\lambda)}_{\tau_1\tau_2}$ can be factorized into linear functions of the following combinations of the anomalous couplings:
\begin{alignat}{2}
\{ & \textbf{T},\,
&& \textbf{S}_{1,\,\gamma} (1+ \Delta g_{1,\,\gamma}) +
\textbf{S}_{1,\,Z} (1+ \Delta g_{1,\,Z}) ,\, \nonumber \\
& \textbf{S}_{2,\,\gamma} \Delta \kappa_{\gamma} +
\textbf{S}_{2,\,Z} (\Delta \kappa_{Z} - \Delta g_{1,\,Z}) ,\,
&& \textbf{S}_{3,\,\gamma} \lambda_{\gamma} +
\textbf{S}_{3,\,Z} \lambda_{Z} ,\, \nonumber \\
& \textbf{S}_{4,\,\gamma} g_{5,\,\gamma} +
\textbf{S}_{4,\,Z} g_{5,\,Z} ,\,
&& i\textbf{S}_{5,\,\gamma} g_{4,\,\gamma} +
i\textbf{S}_{5,\,Z} g_{4,\,Z} ,\, \nonumber \\
& i\textbf{S}_{6,\,\gamma} (\tilde{\kappa}_{\gamma}-\tilde{\lambda}_{\gamma}) +
i\textbf{S}_{6,\,Z} (\tilde{\kappa}_{Z}-\tilde{\lambda}_{Z}) ,\,\,\,
&& i\textbf{S}_{7,\,\gamma} \tilde{\lambda}_{\gamma} +
i\textbf{S}_{7,\,Z} \tilde{\lambda}_{Z} \,\,\, \} \,,
\end{alignat}
where $\textbf{S}$ and $\textbf{T}$ correspond respectively to the $s$ and $t$ channels for $WW$ production, and their helicity indices $\lambda,\,\tau_{1,\,2}$ are not explicitly shown. It is then rather trivial to take the first order derivative of the amplitudes (squared) with respect to the anomalous couplings: To obtain $\left[ \frac{\partial}{\partial \alpha_i} \left( \frac{{\rm d}\sigma}{{\rm d}\Omega_k } \right) \right]_0$ with $\Omega_k = \cos\theta,\, \cos\theta^\ast_{1,\,2},\, \phi^\ast_{1,\,2}$, we just need to implement the simple replacement in the corresponding differential cross sections
\begin{eqnarray}
\left[ F F^\ast \right]_0 &\rightarrow&
\left[ \left( \partial_{\alpha_i}F\right) F^\ast + F \left( \partial_{\alpha_i} F^\ast \right) \right]_0 \nonumber \\
&\rightarrow& \left( \partial_{\alpha_i} F \right) \left[ F^\ast \right]_0 + \left[ F \right]_0 \left( \partial_{\alpha_i} F^\ast \right) \,.
\end{eqnarray}
where for the five most general $C$ and $P$ conserving couplings,
\begin{eqnarray}
\left\{ \begin{matrix}
\partial_{\Delta g_{1,\,Z}} \\
\partial_{\Delta \kappa_{\gamma}} \\
\partial_{\Delta \kappa_{Z}} \\
\partial_{\lambda_{\gamma}} \\
\partial_{\lambda_{Z}}
\end{matrix} \right\} F^{(\ast)} =
\left\{ \begin{matrix}
0 & \textbf{S}_{1,\,Z} & -\textbf{S}_{2,\,Z} & 0 & 0 & 0 & 0 & 0 \\
0 & 0 & \textbf{S}_{2,\,\gamma} & 0 & 0 & 0 & 0 & 0 \\
0 & 0 & \textbf{S}_{2,\,Z} & 0 & 0 & 0 & 0 & 0 \\
0 & 0 & 0 & \textbf{S}_{3,\,\gamma} & 0 & 0 & 0 & 0 \\
0 & 0 & 0 & \textbf{S}_{3,\,Z} & 0 & 0 & 0 & 0
\end{matrix} \right\} \,.
\end{eqnarray}
When the anomalous couplings are correlated by the EW gauge symmetry,
\begin{eqnarray}
\left\{ \begin{matrix}
\partial_{\Delta g_{1,\,Z}} \\
\partial_{\Delta \kappa_{\gamma}} \\
\partial_{\lambda_{\gamma}}
\end{matrix} \right\} F^{(\ast)} =
\left\{ \begin{matrix}
0 & \textbf{S}_{1,\,Z} & 0 & 0 & 0 & 0 & 0 & 0 \\
0 & 0 & \textbf{S}_{2,\,\gamma} - \tan^2\theta_W \textbf{S}_{2,\,Z} & 0 & 0 & 0 & 0 & 0 \\
0 & 0 & 0 & \textbf{S}_{3,\,\gamma} + \textbf{S}_{3,\,Z} & 0 & 0 & 0 & 0
\end{matrix} \right\} \,.
\end{eqnarray}
It is also straightforward to obtain the derivatives with respect to the $C$ or $P$ violating anomalous couplings and generalize it to the second order derivatives, i.e. $\frac{\partial}{\partial \alpha_i \partial \alpha_j} \left( \frac{{\rm d}\sigma}{{\rm d}\Omega_k } \right)$. Now it is trivial to get the linear coefficients $\omega_i$ in Eq.~(\ref{eqn:omegai}),
\begin{eqnarray}
\omega_i (\Omega_k) = \left[ \frac{\partial}{\partial \alpha_i} \left( \frac{{\rm d}\sigma}{{\rm d}\Omega_k } \right) \right]_0
\left( \frac{{\rm d}\sigma_0}{{\rm d}\Omega_k } \right)^{-1} \,.
\end{eqnarray}
As mentioned in section~\ref{sec:distribution}, integrating over the angle $\Omega_k$ can produce the analytical expressions for the coefficients $b_i$ for the total cross section:
\begin{eqnarray}
b_i = \frac{1}{\sigma_0} \int {\rm d} \Omega_k \left[ \frac{\partial}{\partial \alpha_i} \left( \frac{{\rm d}\sigma}{{\rm d}\Omega_k } \right) \right]_0 \,.
\end{eqnarray}

\section{Numerical expressions for the coefficients $\omega_i$}
\label{sec:bi}

In the SM, the numerical expressions for the angular distributions of the $e^+ e^- \rightarrow W^+ W^- \rightarrow f_1 \bar{f}_2 \bar{f}_3 f_4$ process are, at the center-of-mass energy of 240 GeV designed for CEPC, in unit of pb,
\begin{eqnarray}
\frac{{\rm d} \sigma_0}{{\rm d} \cos\theta}
&=& \frac{ 3.420 - 1.496 \cos\theta - 1.026 \cos^2\theta + 0.06429 \cos^3\theta - 0.8394 \cos^4\theta }{ \left( 1-0.9571 \cos\theta \right)^{2} } \,, \nonumber \\
\frac{{\rm d} \sigma_0}{{\rm d} \cos\theta^\ast_{1,\,2}}
&=& 7.440 + 8.155 \cos\theta^\ast_{1,\,2} + 3.449 \cos^2\theta^\ast_{1,\,2} \,, \nonumber \\
\frac{{\rm d} \sigma_0}{{\rm d} \phi^\ast_{1,\,2}}
&=& 2.734 \mp 0.4317 \cos\phi^\ast_{1,\,2} -
0.2080 \cos^2\phi^\ast_{1,\,2} \,.
\end{eqnarray}
For the five $C$ and $P$ conserving aTGCs $\alpha_i = \Delta g_{1,Z}$, $\Delta \kappa_{\gamma,Z}$, $\lambda_{\gamma,Z}$, the linear coefficients $\omega_i (\Omega_k)$ for the differential cross sections are, respectively,
\begin{eqnarray}
\omega_i (\cos\theta) \frac{{\rm d}\sigma_0}{{\rm d}\cos\theta} &=& \frac{1}{ \left( 1-0.9571 \cos\theta \right)^{2}}
\left[ \begin{pmatrix} -0.9030 \\ -2.501 \\ -1.832 \\ -1.342 \\ -0.9832 \end{pmatrix}
+\begin{pmatrix} 1.543 \\ 5.347 \\ 4.325 \\ 3.129 \\ 2.700 \end{pmatrix} \cos\theta \right. \nonumber \\
&& \left. +\begin{pmatrix} -0.7301 \\ -1.669 \\ -1.612 \\ -1.765 \\ -1.683 \end{pmatrix} \cos\theta^2
 +\begin{pmatrix} 1.157 \\ -2.218 \\ -1.625 \\ 0 \\ 0 \end{pmatrix} \cos\theta^3
+\begin{pmatrix} -1.034 \\ 1.062 \\ 0.7776 \\ 0 \\ 0 \end{pmatrix} \cos\theta^4 \right] \,, \nonumber \\
\omega_i (\cos\theta^\ast_{1,\,2}) \frac{{\rm d}\sigma_0}{{\rm d}\cos\theta^\ast_{1,\,2}} &=&
\begin{pmatrix} -0.4215 \\ -2.062 \\ -0.9475 \\ -0.8048 \\ -0.1983 \end{pmatrix}
+\begin{pmatrix} -0.2954 \\ -0.2022 \\ -0.2954 \\ -0.2022 \\ -0.2954 \end{pmatrix} \cos\theta^\ast_{1,\,2}
+\begin{pmatrix} 0.2676 \\ 1.592 \\ 0.7489 \\ 0.1362 \\ 0.1990 \end{pmatrix} \cos^2\theta^\ast_{1,\,2} \,, \nonumber \\
\omega_i (\phi^\ast_{1,\,2}) \frac{{\rm d}\sigma_0}{{\rm d}\phi^\ast_{1,\,2}} &=&
\begin{pmatrix} -0.1058 \\ -0.488 \\ -0.2221 \\ -0.2417 \\ -0.04202 \end{pmatrix}
\pm \begin{pmatrix} 0.6331 \\ 0.2945 \\ 0.8003 \\ 0.2945 \\ 0.8003 \end{pmatrix} \cos\phi^\ast_{1,\,2}
+ \begin{pmatrix} -0.04627 \\ -0.01051 \\ +0.01057 \\ -0.1838 \\ -0.2426 \end{pmatrix} \cos^2\phi^\ast_{1,\,2} \,.
\end{eqnarray}
Using these expressions, it is easy to check that integrations of the distributions $\omega_i$ result in the coefficients $b_i$ for the total cross section:
\begin{eqnarray}
\label{eqn:integration}
\frac{1}{\sigma_0}\int {\rm d} \Omega_k \, \omega_{i} (\Omega_k) \frac{{\rm d}\sigma_0}{{\rm d}\Omega_k} = b_i \,.
\end{eqnarray}

It is phenomenologically more interesting to study the aTGCs in the cases where the physics beyond SM are invariant under the EW gauge symmetry. Under such conditions, the corresponding distributions $\omega$ for $\alpha_i = \Delta g_{1,Z}$, $\Delta \kappa_{\gamma}$, $\lambda_{\gamma}$ are, respectively,
\begin{eqnarray}
\omega_i (\cos\theta) \frac{{\rm d}\sigma_0}{{\rm d}\cos\theta} &=& \frac{ 1}{ \left( 1-0.9571 \cos\theta \right)^{2}}
\left[
\begin{pmatrix} -2.735 \\ -1.977 \\ -2.325 \end{pmatrix}
+\begin{pmatrix} 5.868 \\ 4.111 \\ 5.829 \end{pmatrix} \cos\theta \right. \nonumber \\
&& \left. +\begin{pmatrix} -2.342 \\ -1.208 \\ -3.449 \end{pmatrix} \cos^2\theta
 +\begin{pmatrix} -0.4683 \\ -1.754 \\ 0 \end{pmatrix} \cos^3\theta
+\begin{pmatrix} -0.2559 \\ 0.8393 \\ 0 \end{pmatrix} \cos^4\theta \right] \,, \nonumber \\
\omega_i (\cos\theta^\ast_{1,\,2}) \frac{{\rm d}\sigma_0}{{\rm d}\cos\theta^\ast_{1,\,2}} &=&
\begin{pmatrix} -1.369\\-1.792\\-1.003 \end{pmatrix}
+\begin{pmatrix} -0.5907\\-0.1178\\-0.4976 \end{pmatrix} \cos\theta^\ast_{1,\,2}
+\begin{pmatrix} 1.016\\ 1.378\\0.3352 \end{pmatrix} \cos^2\theta^\ast_{1,\,2} \,, \nonumber \\
\omega_i (\phi^\ast_{1,\,2}) \frac{{\rm d}\sigma_0}{{\rm d}\phi^\ast_{1,\,2}} &=&
\begin{pmatrix} -0.3279\\-0.4241\\-0.2837 \end{pmatrix}
\pm \begin{pmatrix} 1.433\\ 0.06575\\ 1.095 \end{pmatrix} \cos\phi^\ast_{1,\,2}
+ \begin{pmatrix} -0.03570\\-0.01353\\-0.4264 \end{pmatrix} \cos^2\phi^\ast_{1,\,2} \,.
\end{eqnarray}
Analogue to Eq.~(\ref{eqn:integration}), integrations of the $\omega$ distributions lead us to the $b_i$ coefficients in Eq.~(\ref{eqn:bi2}) for the total cross sections.

\end{document}